\documentclass[sigconf,nonacm]{acmart}
\usepackage{url}
\usepackage{amsmath}
\usepackage{amsfonts}
\usepackage{xcolor}
\usepackage{graphicx}
\usepackage{algorithm, algorithmic}
\usepackage{comment}
\usepackage{multicol}
\usepackage{multirow, float, makecell}
\usepackage{caption, subcaption}
\usepackage{tabularray}
\UseTblrLibrary{booktabs}
\usepackage{dsfont}

 \def\cD{{\mathcal{D}}}
\def\cP{{\mathcal{P}}}
\def\cN{{\mathcal{N}}}
\def\RR{{\mathbb{R}}}
\def\ZZ{{\mathbb{Z}}}

\def\R{{\mathcal{R}}}
\def\bv{{\mathbf{v}}}
\def\bu{{\mathbf{u}}}

\def\C{{\mathbb{C}}}
\def\Z{{\mathbb{Z}}}
\def\Ecd{{\mathsf{Ecd}}}
\def\Dcd{{\mathsf{Dcd}}}
\def\Dec{{\mathsf{Dec}}}
\def\StC{{\mathsf{StC}}}
\def\ct{{\mathsf{ct}}}
\def\Mult{{\mathsf{Mult}}}
\def\Add{{\mathsf{Add}}}
\def\Rot{{\mathsf{Rot}}}
\def\EvalMod{{\mathsf{EvalMod}}}
\def\slot{\mathrm{slot}}
\def\coeff{\mathrm{coeff}}
\def\sk{{\mathsf{sk}}}
\def\relk{{\mathsf{relk}}}
\def\rotk{{\mathsf{rotk}}}
\def\RingPack{{\mathsf{RingPack}}}
\def\Re{{\mathrm{Re}}}
\def\Im{{\mathrm{Im}}}
\def\RS{\mathsf{Rescale}}
\def\BTS{{\mathsf{BTS}}}

\def\qry{\mathsf{qry}}
\def\db{\mathsf{db}}
\def\MQ{{\mathbf{M}_{\qry}}}
\def\MDB{{\mathbf{M}_{\db}}}
\def\NDB{{N_{\mathsf{db}}}}
\def\NBQ{{N_{\mathsf{qry}}}}

\def\hd{{\mathrm{hd}}}
\def\dist{{\mathrm{dist}}}

\def\cl{{\mathrm{cl}}}

\def\toci{{\mathsf{ToCI}}}
\def\fromci{{\mathsf{FromCI}}}
\def\ci{{\mathsf{CI}}}
\def\pk{{\mathsf{pk}}}

\def\sh{{\mathsf{sh}}}
\def\evk{{\mathsf{evk}}}
\def\CCMM{{\mathsf{CCMM}}}

\def\CtS{{\mathsf{CtS}}}
\def\dnum{{\mathsf{dnum}}}
\def\db{{\mathrm{db}}}
\def\ctres{{\mathsf{ct_{\mathrm{out}}}}}
\def\ecomp{e_{\mathrm{comp}}}

\def\ebts{e_{\mathrm{bts}}}
\def\pre{\mathsf{pre}}

\def\bc{{\mathbf{c}}}
\def\bm{{\mathbf{m}}}
\def\bu{{\mathbf{u}}}
\def\bv{{\mathbf{v}}}

\title{Private Iris Recognition with High-Performance FHE}

\author{Jincheol Ha}
\authornote{Corresponding author.}
\affiliation{
\institution{Cryptolab, Inc.}
\city{Seoul}
\country{Republic of Korea}
}
\author{Guillaume Hanrot}
\affiliation{
\institution{Cryptolab, Inc.}
\city{Lyon}
\country{France}
}
\author{Taeyeong Noh}
\affiliation{
\institution{Cryptolab, Inc.}
\city{Seoul}
\country{Republic of Korea}
}
\author{Jung Hee Cheon}
\affiliation{\institution{Cryptolab, Inc.} \institution{Seoul National University}
\city{Seoul}
\country{Republic of Korea}
}
\author{Jung Woo Kim}
\affiliation{
\institution{Cryptolab, Inc.}
\city{Seoul}
\country{Republic of Korea}
}
\author{Damien Stehl\'e}
\affiliation{
\institution{Cryptolab, Inc.}
\city{Lyon}
\country{France}
}

\begin{abstract}
Among biometric verification systems, irises stand out because they offer both convenient capture and high accuracy even in large-scale databases of billions of users. For example, the World~ID project aims to provide authentication to all humans via iris recognition, with millions already registered. Storing such biometric data raises privacy concerns, which can be addressed using privacy-enhancing cryptographic techniques. 

To protect the iris codes of the database and queriers, Bloemen \emph{et al.} [IACR~eprint~2024/705] describe a solution based on 2-out-of-3 Secret-Sharing Multiparty Computation (SS-MPC), for the World~ID setup. The protocol takes as input iris codes of queriers and outputs bits telling if each of these codes matches a code stored in the database. In terms of security, unless an adversary corrupts 2~servers, the iris codes remain confidential and nothing leaks beyond the result of the computation. Their solution is able to match~$32$
users against a database of~$2^{22}$ iris codes in~$\approx 2$s , using~24 H100 GPUs, more than 40~communication rounds and $81$GB/party
of data transferred (the timing assumes a network speed above~3Tb/s). 

The purpose of this work is to explore the use of Threshold Fully Homomorphic Encryption (ThFHE) for the same task using the same approach, thereby providing a comparison of two leading privacy-preserving technologies in the biometric context from the perspectives of efficiency, security, and deployment.

ThFHE enables arbitrary computations on encrypted data, while secret-sharing the decryption key: a given number~$t$ of decryptors among~$n$ users must collaborate to obtain the output of the computation, while~$t-1$ users cannot learn anything that is not already public. The ThFHE solution brings a number of security advantages: no trusted setup, the encrypted database and queries can be public, the parameters~$t$ and~$n$ can be increased and active security can be added without significant performance degradation.  

Our proof-of-concept implementation of the computation phase handles $32$~eyes against a database of $7\cdot 2^{14}$ iris codes, ensures $\mathsf{FA}$ (false acceptance rate) and~$\mathsf{FR}$ (false rejection rate) 
within $10^{-9}$ from the values of the underlying iris recognition system, and requires~$\approx 1.8$s computation time, 
using 8 RTX-5090 GPUs. To this, one should add~2 to 3 rounds of communication (depending on deployment choices).
This gives a similar time and cost performance as the SS-MPC solution, but the saving in communication allows to maintain the performance without: 1)~assuming an ultra-high-bandwidth network between non-colluding parties and 2)~restricting to 2-out-of-3 security. 
We perform the matching using the CKKS (Th)FHE scheme. Our main technical ingredients are the use of recent progress on FHE-based linear algebra boosted using int8 GPU operations, and the introduction of a folding technique to reduce the number of ciphertexts to be processed as early as possible. 
\end{abstract}
 
\setcopyright{none}
\renewcommand\footnotetextcopyrightpermission[1]{}

\begin{document}

\maketitle

\section{Introduction}
Accurate biometric recognition solutions   compare a query against a database of stored values that characterize legitimate users (this is often referred to as the $1:N$ scenario). This raises major privacy concerns: the deployment of cryptographically secure solutions is necessary 
not only to store such databases but also to compute on them in a privacy-preserving manner. The query deserves the same level of protection. 

Among various biometric primitives (face, fingerprint, heart rhythm, etc) we choose to illustrate our approach with an iris-based biometric solution. Iris recognition offers strong accuracy even at large scale; this explains the choice of World~ID to rely on it 
to design an authentication system for all humans. In this context, Bloemen et al.~\cite{BGKSW24} use secret-sharing multiparty computation (SS-MPC) for secure storage and recognition. More concretely, they use a protocol based on 2-out-of-3 secret sharing: 3 parties store the database shares, and 2 out of them must participate in the computation (the third party creates redundancy in case a party becomes unavailable); in terms of security, a single party has no information on the underlying data, but 2 colluding parties can reconstruct everything. Bloemen et al.~\cite{BGKSW24} proposed a GPU-based implementation with impressive large-scale performance. 

A comparison between the SS-MPC approach from~\cite{BGKSW24} and fully homomorphic encryption (FHE) techniques is outlined in a blog entry~\cite{WK24}.
It rules out FHE in this context by pinpointing fundamental weaknesses in FHE architectures. Notably,  FHE security is claimed to rely on a no-collusion assumption between the database holder and the key holder. FHE's computational cost is claimed to be prohibitive, with 17ms required for a mere addition of 16-bit integers on GPU. Finally, the communication costs and database storage footprint are claimed to be massive (37MB to transmit a query corresponding to a single  iris code, 3.5TB to store a database of 100000 iris codes), making the deployment of an FHE-based solution seemingly unrealistic.   

Symmetrically, SS-MPC solutions, including~\cite{BGKSW24}, suffer from several drawbacks. First, the cost scales poorly with the $t$-out-of-$n$ access pattern. Large values of~$t$ and~$n$ are desirable for security and robustness: a recent voting failure at the IACR showed that $3$-out-of-$3$ secret-sharing is not robust;\footnote{See \url{https://www.nytimes.com/2025/11/21/world/cryptography-group-lost-election-results.html}} further, $2$-out-of-$3$ secret-sharing does not provide security against collusions. Second, the side-channel attack surface is large,  as any party is computing on sensitive data. Third, such solutions are latency-bound, as they often require tens of communication rounds: Bloemen et al.~\cite{BGKSW24} assume an NVLink between the parties, which means that they must be at same physical location, putting more stress on the no-collusion assumption of 2-out-of-3 secret sharing. 

In the present work, we explore the use of threshold fully homomorphic encryption (ThFHE)~\cite{AJLA12,BGG+18} as an alternative to SS-MPC for large-scale iris recognition. ThFHE is an extension of FHE where the decryption key is secret-shared: ThFHE involves secret-sharing aspects for key generation and decryption, but is otherwise identical to FHE.  
Importantly, ThFHE does not suffer from the previous security limitations: the main challenge is to tame its computation cost.

\subsection{Contributions}

We design and optimize a high-performance (Th)FHE protocol for iris recognition, and compare performance and deployment constraints to the recent solution from~\cite{BGKSW24} based on SS-MPC. To allow accurate comparison, we use the same iris recognition process, based on binary iris codes and Hamming distance~\cite{Dau04}. Our method extends to modern cosine-similarity based iris recognition models, such as~\cite{RWWST19,YXF21,WHWHS22,NFSR23,SWWWS25} by redesigning the FHE parameters.

Our proof-of-concept implementation shows that using a cluster of 8 RTX-5090 GPUs gives a match of a batch of 32 query eyes against a database of size $7\cdot 2^{14}$ in under 2s. The database storage corresponding to 100k entries is $\approx 123$GB (30$\times$ smaller than estimated in~\cite{WK24}), while the query size of a single iris code is $512$KB ($74 \times$ smaller than estimated in~\cite{WK24}).

The solution scales to larger databases by adding computational resources. This performance is comparable to that of~\cite{BGKSW24} when taking into account the relative prices of the considered GPUs. 
We argue, however, that our solution has better scaling properties when one wants to further distribute security: our computation cost is almost agnostic to the number~$n$ of parties  involved and the threshold~$t$. Further, the SS-MPC approach requires a very large amount of communication (tens of~GB) and communication rounds (around~40) requiring ultra-high-bandwidth network for efficient deployment, while ours has low communication requirements (hundreds of~KB per party) and can be deployed over WAN.

Achieving this performance requires a fast FHE library (in our case, HEaaN2~\cite{heaanlib2}, implementing the CKKS scheme~\cite{ckks}), and, crucially, the design of efficient FHE evaluation algorithms for the task at hand. The core of the iris recognition process consists in two main steps: a first step matching query data against all database entries to obtain scores with a matrix multiplication, and a second step comparing the scores to a cutoff value. 

We design two homomorphic instantiations of the core computation. The first one attempts at minimizing the cost of score computation, but incurs large bootstrapping costs that dominate the overall cost. Bootstrapping~\cite{CHKKS18} is a ciphertext maintenance operation that allows to continue homomorphic computations. The second one explores the idea of increasing the linear algebra costs in order to compute with scores before bootstrapping them. It leverages the ability to compute on scores to \emph{fold} them before bootstrapping: after a small-degree polynomial evaluation, they are added by batches of size~$k$ (in our implementation, we take~$k=16$) and then the sums are  bootstrapped; for a large number of ciphertexts, this reduces the bootstrapping cost by a factor~$k$. 

In our implementation, we considered the very recent encrypted matrix algorithm from~\cite{BCHPS25} that reduces the task to plaintext matrix multiplication. We optimized it by taking advantage of int8 tensor cores and the NVIDIA cuBLAS library. 

Finally, we describe an integration of our algorithms within a larger end-to-end system for privacy-preserving iris recognition using $t$-out-of-$n$ ThFHE, providing high security guarantees under mild communication assumptions. This includes preprocessing the query  to allow its transmission with small communication costs, and post-processing the results  to enable secure threshold decryption.

\subsection{Technical overview}
We consider the core of the iris recognition process. It takes as input a database of iris templates with $n_{\db}$ entries, and a query consisting in $\rho$ query vectors; the use of multiple query vectors (``rotations") to encode a single query allows to compensate for physical capture conditions. 
A large number of matching scores is obtained as normalized inner products of each query vector with each database entry. Each of those scores is then compared with a cutoff value, set depending on the properties of the underlying encoding. The authentication of the query user is successful if one of the resulting comparisons is positive. 

\noindent
\textbf{First approach.} We consider a first FHE-based approach directly following the above description. In order to minimize the linear algebra cost of the first step, we perform this computation at the smallest possible ciphertext modulus. The resulting scores are then bootstrapped so that the subsequent comparison with the cutoff can be performed homomorphically. This comparison produces~$\rho \cdot n_{\db}$ matching results per query eye. 
Prior to recombining those results, we need to 
ensure that they lie in~$\{0, 1\}$, a property which is difficult to guarantee using CKKS-based comparison when values are very close to the cutoff. We perform a \emph{discretization} step inspired from~\cite{KN24} that forces the output to be in~$\{0, 1\}$, allowing us to group all~$\rho \cdot n_{\db}$ matching results  into a single bit via an OR-tree (such binary computations can be performed in CKKS using~\cite{DMPS24,BCKS24}).

\noindent
\textbf{Folding.} When handling a very large database, the main drawback of the first approach is the need to bootstrap all score values right after the score computation. This may correspond to a huge number of ciphertexts.  In order to reduce the bootstrapping cost, we introduce a pretreatment of the score. This pretreatment consists in evaluating a polynomial of the score values, sending non-matching scores to small values and matching scores to large values. We build score batches of size~$k$ that contain at most one matching score; the pretreated scores in one batch can then be added while preserving the relevant information, i.e., whether this batch contains a matching score or not. This reduces the bootstrapping cost of the first method by a factor~$k$, while increasing only moderately the pre-bootstrapping costs. We combine this folding idea with the other components of the first approach to obtain a complete chain. 

\noindent
\textbf{System integration.} We consider the integration of this core matching evaluation within a full system performing $1:N$ iris recognition. 
Our folding technique leads to a rather large ciphertext modulus for the query, implying a large expansion factor in communication. We describe a query format minimizing this communication cost, and a bootstrapping-based query preprocessing  converting the format to the one required by the score computation. 
Regarding security, we rely on $t$-out-of-$n$ ThFHE. In order to allow for secure partial decryption, we complete the core computation chain by a high precision cleaning step~\cite{KSS25}.

\subsection{In praise of ThFHE}

The iris recognition process based on ThFHE has several advantages 
compared to solutions based on SS-MPC, and addresses the FHE security concerns raised by~\cite{WK24}. 
\begin{itemize}
\item[$\bullet$] Most of the work, the homomorphic matching, is done on encrypted data and can hence be \emph{performed publicly}. In particular, the encrypted database can be made public.  It implies that the load can be arbitrarily distributed between computing servers. The need to trust the servers can also be decreased by adding redundancy: the computation is public and deterministic and can hence be reproduced. There is also no need to add physical protection to the  servers.   
\item[$\bullet$] The ThFHE protocol is \emph{communication-light}. With SS-MPC, if the computing servers secret-share sensitive data and should not collude, one would prefer that they are in different physical locations, which increases communication costs. In the ThFHE protocol, there is no communication in the heavy step, unless the work is distributed on different computing servers: even in that case, as it is useless for them to collude, they can be at the same location, hence decreasing communication costs. 
\item[$\bullet$] \emph{Efficiency  scales well} with the number of decryptors and with an upgrade to active security: indeed, only the rather light decryption phase is impacted by an increase of the number~$n$ of users and decryption threshold~$t$.  For active security, only the partial decryptions run by the decryptors need to be strengthened by zero-knowledge proofs, the main computation remaining the same.
\end{itemize}

\subsection{Related works}
Most existing works on biometrics secured by homomorphic encryption focus on the $1:1$ authentication scenario, where the user~ID is provided together with the template vector, so that matching is performed against a single database entry. In this context, many papers~\cite{CCKL16,KPR20,KPR21, KPVR21,ALGRB23,MSD23,BZCRK023,AB24,BHVP24,PM24,LGCN25,WOAS25}  focus on the secure score computation and perform the comparison with  the cutoff in clear, after decryption, thus disclosing some information regarding the protected data; FHE may also be combined with MPC to securely perform the comparison~\cite{BHVP24}. 
The best of those works typically demonstrate strong efficiency for $1:1$ score computation with timings of the order of tens of milliseconds on standard CPUs, for medium template dimensions (128-512). 

Some recent works explore the $1:N$ setting, while focusing only on score computation. In~\cite{EJB22}, the authors 
report a performance of $\approx 0.03$ms per score on a 10-core CPU  with template dimension~128, at the expense of a high query communication cost. The authors of~\cite{CKSWK24}, on the other hand, use a compact query format 
and report a cost of $\approx 0.07$ms per score computation with template dimension~$128$, on 8 CPU cores. Finally, FHE is combined with functional secret sharing in~\cite{IKCDO24} to perform a secure comparison to the cutoff; score computation amounts to $\approx$ 0.15ms per score for template dimension $128$, on a 4-core CPU.

In addition to $1:N$ score computation, the authors of~\cite{ICDO23} also handle the comparison step homomorphically. To scale to medium-sized databases, they introduce a group testing idea which bears resemblance to our \emph{folding} approach, by computing a somewhat heuristic approximation of the maximum of a number of scores; the heuristic nature of the approximation leads to a rather high rate of false rejections. Using somewhat homomorphic encryption, they process $2^{14}$ scores in template dimension 128 in $\approx 9.4s$ on one CPU. 
 \section{Preliminaries}

We use boldface letters for matrices and vectors. We let the integer ring be denoted by ${\mathbb Z}$ and the field of real (resp.\ complex) numbers by~${\mathbb R}$ (resp.\ ${\mathbb C}$). The sets ${\mathbb R}^n$ and ${\mathbb C}^n$ are equipped with their coordinate-wise ring structure, with multiplication denoted by~$\odot$.
We let $\|\cdot \|_1$ denote the $L^1$-norm, i.e., $\|{\bf x}\|_1 = \sum_{i=0}^{n-1} |x_i|$ for ${\bf x}\in \RR^n$. 
For $x\in \RR$, we define $\lfloor x\rceil = \lfloor x+1/2\rfloor$, where~$\lfloor \cdot \rfloor$ refers to the integer part. The notation~$\log$ refers to the base-2 logarithm. Finally, we use~$\wedge$,  $\vee$ and~$\oplus$ for the and, or and xor boolean operators, respectively, and let them operate coordinate-wise on  vectors. 

\subsection{Plain iris recognition}
\label{sse:plain}
\noindent
In this section, we summarize the iris recognition process proposed by Daugman~\cite{Dau04} used in the SS-MPC solution~\cite{BGKSW24}.
The first step is image segmentation, converting an input iris image into two binary vectors of the same size: an \emph{iris code} $\mathbf{c} \in \{0, 1\}^\ell$ obtained by segmenting and normalizing the input iris image, and a \emph{mask} $\mathbf{m} \in \{0, 1\}^\ell$ used to occlude non-iris parts of $\mathbf{c}$ such as eyelids and eyelashes in the iris image.
A pair of an iris code and a corresponding mask is called \emph{iris template}.
Daugman used 2-dimensional Gabor wavelets for this iris image processing, and the World ID infrastructure used machine learning techniques for more precise image processing.
We omit the details for this part and assume an iris template $(\mathbf{c}, \mathbf{m})$ is given as input.

\subsubsection{Distance on iris templates}
Given an iris template $(\mathbf{c}, \mathbf{m})$, the usable iris part from the original iris image corresponds to $\mathbf{c} \land \mathbf{m} \in \{0, 1\}^\ell$.
For a pair of iris templates $((\mathbf{c}_1, \mathbf{m}_1), (\mathbf{c}_2, \mathbf{m}_2))$, the componentwise AND $\mathbf{m}_1 \land \mathbf{m}_2$ denotes the overlapped iris area. The distance between two iris templates is defined as: 
\begin{align}
    \dist\left((\mathbf{c}_1, \mathbf{m}_1), (\mathbf{c}_2, \mathbf{m}_2)\right)
    & = \frac{\hd(\mathbf{c}_1 \land \mathbf{m}_1, \mathbf{c}_2 \land \mathbf{m}_2)}{\Vert \mathbf{m}_1 \land \mathbf{m}_2 \Vert_1}
    \\
    & = \frac{\langle \mathbf{c}_1 \oplus \mathbf{c}_2, \mathbf{m}_1 \land \mathbf{m}_2 \rangle}{\Vert \mathbf{m}_1 \land \mathbf{m}_2 \Vert_1} \nonumber
\end{align}
where the Hamming distance $\hd$ between two binary vectors~$\mathbf{x}$ and~$\mathbf{y}$ is defined by $\hd(\mathbf{x}, \mathbf{y}) = \Vert \mathbf{x} \oplus \mathbf{y} \Vert_1$.

If either iris code $\mathbf{c}_1$ or $\mathbf{c}_2$ is a uniform binary vector independent of the other one, then $\mathbf{c}_1 \oplus \mathbf{c}_2$ is expected to behave as a uniform binary vector, so that the distance is close to~$1/2$.
Daugman~\cite{Dau04} found that two eyes of different individuals and also different eyes of the same individual have independent behavior,  obtaining a close-to-normal distribution centered near~$1/2$ for the distance between the iris templates.
Conversely, if the iris templates are obtained from the same eye so that they are close to each other, then the distance would be close to $0$.
By using this difference between the matching case (i.e., iris templates from the same eye) and the non-matching case, we can decide whether two iris templates come from the same eye or not.
But even for the same eye, obtaining close iris templates is not easy in the real world because we cannot ideally reproduce the same environment when taking the iris images.

\subsubsection{Iris template rotations}
To mitigate the  issue above, one of the iris templates to be matched is rotated (both code and mask) by a small offset several times, and the distance to the other one is computed independently for all rotations.
If the irises are independent, the rotations do not affect the distribution of the distance (Daugman experimentally showed that they can be assumed independent).
When considering the same eye, it can however happen that at least one of the rotations would result in a small distance (the distance from the other rotations may give a larger distance as in the non-matching case).
Hence, by taking the minimum distance among different rotations, one can compensate for the non-ideal physical measurement of the iris image.

We can choose a cutoff $t \in (0,1)$ and decide that two iris codes come from the same eye when the minimum distance among rotated iris codes is smaller than~$t$.
A larger~$t$ is more likely to result in incorrect matching (false positive), while a smaller~$t$ is more likely to result in incorrect non-matching (false negative).

\subsubsection{Masked bitvector representation}\label{sec: masked bitvector}
To compute the distance more efficiently, Bloemen et al.~\cite{BGKSW24} adopted a masked bitvector representation as follows.
Given two iris data $(\mathbf{c}_i, \mathbf{m}_i) \in \{0, 1\}^{\ell}$ for~$i \in \{1,2\}$, define a ternary vector $\mathbf{c}_i'$ by
\[
    \mathbf{c}_i' = \mathbf{m}_i - 2 \cdot (\mathbf{c}_i \land \mathbf{m}_i) \in \{-1, 0, 1\}^\ell \enspace.
\]
Then, we have
\(
    2 \cdot \langle \mathbf{c}_1 \oplus \mathbf{c}_2, \mathbf{m}_1 \land \mathbf{m}_2 \rangle = \Vert \mathbf{m}_1 \land \mathbf{m}_2 \Vert_1 - \langle \mathbf{c}_1', \mathbf{c}_2' \rangle,
\)
so that we can replace the matching decision criteria by
\begin{equation}
    \label{eq:masked bitvector}
    \frac{\langle \mathbf{c}_1', \mathbf{c}_2' \rangle}{\Vert \mathbf{m}_1 \land \mathbf{m}_2 \Vert_1} > 1 - 2t\enspace.
\end{equation}
In this work, we call the left side of~\eqref{eq:masked bitvector}  the \emph{score} of the iris pair.

\subsubsection{Homomorphic context and database matching}
In the context of homomorphic evaluation, implementing a decision criterion like~\eqref{eq:masked bitvector} is not  feasible efficiently. Rather, we assume that we are given two intervals $\cN = [s,t_0]$ and $\cP=[t_1,u]$ with~$t_0 < t_1$ such that all scores belong to~$[s, u]$, and we implement the following:
\begin{eqnarray}
\frac{\langle \mathbf{c}_1', \mathbf{c}_2' \rangle}{\Vert \mathbf{m}_1 \land \mathbf{m}_2 \Vert_1}  \in  \cN  &  \Rightarrow  & \mbox{Return \texttt{false}}\enspace,  \nonumber \\
    & & \label{eq:condition}\\[-0.3cm]
    \frac{\langle \mathbf{c}_1', \mathbf{c}_2' \rangle}{\Vert \mathbf{m}_1 \land \mathbf{m}_2 \Vert_1}  \in \cP & \Rightarrow & \mbox {Return \texttt{true}} \nonumber \enspace.
\end{eqnarray}

In the case where the score is in $(t_0, t_1)$, we consider the output to be unspecified but it should be either \texttt{true} or \texttt{false}, 
to allow further computations that assume the result is binary. We let the result of this operation be denoted by $\mathsf{match}((\bc_1, \bm_1), (\bc_2, \bm_2))$.

For the  matching of a set of iris templates for rotations of a fixed iris $\mathbf{C}^{(\qry)} = (\mathbf{c}_r^{(\qry)}, \mathbf{m}_r^{(\qry)})_{0 \le r < \rho}$ against a whole database of iris templates $\mathbf{C}^{(\db)} = (\mathbf{c}^{(\db)}_i, \mathbf{m}^{(\db)}_i)_{0\le i < n_\db}$, we 
define:
\[
\mathsf{match}_{\db}(\mathbf{C}^{(\qry)}, \mathbf{C}^{(\db)})
= \hspace*{-.3cm} \bigvee_{\substack{0 \leq i< n_\db \\ 0 \leq r < \rho}} \hspace*{-.3cm} \mathsf{match}\left((\mathbf{c}^{(q)}_r\hspace*{-.1cm}, \mathbf{m}^{(q)}_r), (\mathbf{c}^{(\db)}_i\hspace*{-.1cm}, \mathbf{m}^{(\db)}_i)\right) \enspace. 
\]

\subsubsection{Modern iris recognition models}
\label{sec:modern ir models}
We note that the homomorphic context remains the same in the modern iris recognition models based on cosine similarity~\cite{RWWST19,YXF21,WHWHS22,NFSR23,SWWWS25};
they extract real-valued iris code vectors and compute inner product between them to get the distance.
Furthermore, several models~\cite{RWWST19,WHWHS22,SWWWS25} are mask-free, reducing normalization cost and preventing information leakage from the iris mask~\cite{KBB17}.

\subsection{CKKS}
CKKS  handles complex numbers as cleartexts. It is, by construction, an approximate system, providing operations on cleartexts  with a precision that can be parameterized. 

Below, we provide a brief description of elementary CKKS operations. More CKKS background, on bootstrapping, ring-packing and the use of the maximal real subfield is provided in Appendix~\ref{app:moreCKKS}.
 
\subsubsection{Encoding and decoding}
Let $N$ be a power-of-two integer. The message ring is $(\C^{N/2}, +, \odot)$, and the plaintext space is the 
ring $\R = \Z[X]/(X^N+1)$. We let $\zeta = \exp(I\pi/N)$ be a primitive $2N$-th root of unity. CKKS uses two encoding maps $\Ecd_{\slot}, \Ecd_{\coeff}: (x_i)_{0\le i < N/2} \mapsto \lfloor \Delta P\rceil$ for a scaling factor~$\Delta$ and where~$P$ is respectively defined by $P(\zeta^{5^i}) = x_i$ for  $0\le i < N/2$, and $P = \sum_{i=0}^{N/2-1} (\Re(x_i) + \Im(x_i) X^{N/2}) \cdot X^i$.
The scaling factor~$\Delta$ controls the level of discretization: the larger~$\Delta$, the higher the precision. 
Through CKKS encoding the arithmetic of the plaintext space corresponds to the arithmetic of the message ring. In the sequel, we shall write $\Ecd$ for $\Ecd_{\slot}$ and $\Dcd$ (resp.\ $\Dcd_{\coeff}$) for the inverse map of $\Ecd$ (resp.\ $\Ecd_{\coeff}$), up to rounding. In the case of $\Ecd_{\slot}$, each coordinate of the message vector is called a slot.

\subsubsection{Ciphertexts}
CKKS relies on the  RLWE problem~\cite{SSTX09,LPR10}. Its secret key~$\sk$ lies in~$\R$ and its public key and ciphertexts are pairs $\ct = (a, b)$ of elements in $\R_Q = \Z_Q[X]/(X^N+1)$. Decryption is defined as~$\Dec_\sk((a,b))=a\cdot \sk + b$,
whereas encryption takes as input a plaintext~$P$ and a public key, and returns a ciphertext~$(a,b)$ such that~$a \cdot \sk + b \approx P$. We refer to~\cite{ckks} for key generation and the description of encryption.

\subsubsection{Operations}
CKKS provides the following operations: 
\begin{itemize}
    \item $\Add$. On input $\ct = (a,b), \ct'=(a',b')$, return $(a+a', b+b')$.  \item $\Mult$. On input $\ct = (a,b)$, $\ct'=(a', b')$, and a relinearization key $\relk$, return a ciphertext $\Mult(ct, ct')$ that decrypts then decodes to $\approx  \Dcd(\Dec(\ct)) \odot \Dcd(\Dec(\ct'))$. 
    \item $\Rot$. On input $\ct = (a,b)$ encrypting $(x_i)_{0\le i < N/2}$, an integer $0\le \ell < N/2$ and an associated key $\rotk_\ell$, return 
    a ciphertext $\Rot_{\ell, \rotk_\ell}(\ct)$ that decrypts then decodes to
    $\approx (x_{i+\ell \bmod N/2})_{0\le i < N/2}$.
\end{itemize}

Based on  addition and multiplication, one can perform slot-wise polynomial evaluation. The cost of the  Paterson-Stockmeyer algorithm~\cite{PaSt73} is dominated by $O(\sqrt{d})$ ciphertext-ciphertext multiplications, where~$d$ is the polynomial degree.

\subsubsection{Maintenance}
When multiplying two encoded plaintext with scaling factor $\Delta$, one obtains a plaintext encoded with scaling factor~$\Delta^2$. In order to preserve the scaling factor, a division by~$\Delta$ is required. This is realized, for $(a,b)\in R_Q^2$ and~$\Delta | Q$, as
\[
\RS_\Delta((a, b)) = 
\left(\left\lfloor \frac{a}{\Delta} \right\rceil, 
\left\lfloor \frac{b}{\Delta} \right\rceil \right)
\bmod \frac{Q}{\Delta} \enspace. 
\]
In practice, one uses $\RS_{q}$ for $q\approx\Delta$, so that different~$q$'s may be used for the same~$\Delta$ while allowing RNS arithmetic for ciphertexts.

$\RS$ leads to a reduction of the ciphertext modulus: during a computation, the modulus decreases with multiplications. This reduction is proportional to~$\Delta$, which controls the precision of the computation: modulus loss is directly related to precision needs. In a simplified view, one can assume that modulus can only take a finite number of values $Q_{L} > Q_{L-1}> \ldots> Q_0$; in that case, we refer to the \emph{level} of a ciphertext in~$[0, L]$ rather than to its modulus. 

When modulus becomes low, the ciphertext must be refreshed via  bootstrapping ($\BTS$)~\cite{CHKKS18}. Homomorphic algorithm design targets a sharp control of multiplicative depth and plaintext precision.

\subsection{Linear algebra in CKKS}\label{sse:ccmm}
The start of the iris recognition computation is a matrix-matrix product, 
where both operands are encrypted (CCMM). 
We focus on the RGSW-based approach described in~\cite{BCHPS25}. We summarize here the properties required for the description of our contributions. We make extensive use of the MSRLWE (also known as ``shared-a") ciphertext format as in~\cite{BCHPS24,BCHPS25}; in the case of large matrices, using this format reduces both the memory footprint and the computational cost. 
When used to multiply a large preprocessed database~$\MDB$ of dimension $d_1\times d_2$ by a query~$\MQ$ of dimensions $d_2 \times d_3$, using scaling factor $\Delta$, and ring degrees~$\NDB$ and~$\NBQ$ for the encryption layer of~$\MDB$ and~$\MQ$:
\begin{itemize}
\item the encrypted database $\MDB$ is represented by four matrices modulo an integer $\approx Q^2/\Delta$ of respective dimensions $\NDB\times d_2$, $d_1 \times d_2$, $\NDB\times \NBQ$ and $d_1 \times \NBQ$; here, we complete the database with zero entries so that $\NDB|d_1$, and pad each database entry with zeroes so that $\NBQ | d_2$; 
\item the encrypted query $\MQ$ is represented by two matrices modulo $Q$ of respective dimensions $\NBQ\times d_3$ and $d_2 \times d_3$; 
\item the encrypted result $\MDB \cdot \MQ$ is represented by two matrices modulo an integer~$\approx Q/\Delta$ of dimensions $\NDB\times d_3$ and $d_1\times d_3$; the second matrix corresponds to $d_1d_3/\NDB$ ciphertexts with ring-degree~$\NDB$ and modulus $\approx Q/\Delta$; each ciphertext encrypts~$\NDB$ consecutive coefficients of one column of~$\MDB\cdot \MQ$;
\item the computation is dominated by four plaintext matrix-matrix products modulo an integer $\approx Q^2/\Delta$. 
\end{itemize}

We let this algorithm be denoted by $\CCMM(\widehat{\MDB}, \widehat{\MQ})$, where $\widehat{\MDB}$ and~$\widehat{\MQ}$ are the encryptions of~$\MDB$ and~$\MQ$, respectively. 
It was designed to multiply encrypted matrices with coefficient-encoded messages,
and return an encrypted matrix with coefficient-encoded messages. It is however compatible with~$\MDB$ using slot-encoding and~$\MQ$ using coefficient-encoding. A similar situation is discussed in~\cite{BCHPS24}. Conversely, the algorithm is also able to handle a slot-encoded~$\MDB$ by replacing $\MDB$ by $\MDB \cdot {\bf M}_\CtS$, where~${\bf M}_{\CtS}$ is the $\CtS$ matrix for ring degree~$\NBQ$~(see~\cite{CHKKS18}). In all cases, the encoding type of the result is that of~$\MDB$.

\subsection{Two-way classification in CKKS}\label{sec:clean}
Let $I_0, I_1$ be two disjoint intervals and~$\varepsilon>0$. We say that $\cl: \mathbb{R}\rightarrow \mathbb{R}$ is an $(I_0, I_1, \varepsilon)$-classification function if $|\cl(x) - b| \le \varepsilon$ when~$x \in I_b$, for $b \in \{0, 1\}$; the behavior of~$\cl$ outside of $I_0 \cup I_1$ is unspecified, but in this paper we only consider realizations such that~$\cl(x) \in [-\varepsilon, 1+\varepsilon]$ for~$x$ in the convex hull of~$I_0 \cup I_1$.
We use this classification primitive for implementing~\eqref{eq:condition}
and for making imprecise approximations to binary values $\{0, 1\}$ more precise (both for correctness and for threshold decryption security). 
In the sequel, we write $\textsf{Classify}_\varepsilon(I_0, I_1, \ct)$ for the homomorphic evaluation of an $(I_0, I_1, \varepsilon)$-classification function on entry~$\ct$. For very small~$\varepsilon$, we make use of the solution described in~\cite[Figure~2]{KSS25}, to lower modulus usage towards the end of the computation. Borrowing from the terminology introduced in~\cite{DMPS24}, we shall call \emph{cleaning} this specific classification regime.

We refer to Appendix~\ref{app:classification} for more details on the implementation of classification in CKKS. \section{A first algorithm} \label{sec:method1}

For the sake of simplicity, throughout Sections~\ref{sec:method1} and~\ref{se:method2}, we assume that the embedding dimension~$d$ of the iris codes coincides with the CKKS ring degree~$N$, and that the latter is the same for all ciphertexts; this is not required for the implementation described in Section~\ref{sec:implem}.

We now describe a first approach to homomorphically evaluate the iris recognition algorithm.
We assume that we are provided  the  CKKS-encrypted database  of the iris codes $(\bc_k^{(\db)})_{0\le k < n_{db}}$ of the registered users.
When the encrypted query is received, the goal is to assess whether the query eye is registered in the database or not: this corresponds to the evaluation of the \textsf{match}$_{\textrm{DB}}$ primitive defined in Section~\ref{sse:plain}.
We consider the following steps.
\begin{enumerate}
    \item {[Query preprocessing]} Given an encrypted query in a format minimizing communication, turn the query into a suitable format for score computation. 
    \item {[Batched inner products computation]} Compute the inner products by CCMM; these are the numerators of the scores. 
\item {[Normalize]} Divide the inner product outputs by the denominators to obtain the scores.\label{step:normalize}
    \item{[Classify]} Turn the scores in the union of intervals $\cN \cup \cP$ into values in~$\{0, 1\}$ by evaluating a classification function; scores outside $\cN \cup \cP$ are sent to $(0, 1)$. \label{step:classify}
    \item{[Result post-processing]} Send values in $(0, 1)$ to~$\{0, 1\}$  and group them
into a single bit.
\end{enumerate}

Step~\ref{step:normalize} is performed using plaintext-ciphertext multiplications, where the plaintexts store the inverses of the denominators.
We assume the iris masks are given in plaintexts to avoid heavy homomorphic division. Alternatively, one could use mask-free models (see Section~\ref{sec:modern ir models}).

Step~\ref{step:classify} is handled using the strategy outlined in Appendix~\ref{app:classification}. 
We describe the details of the other steps in the following subsections. The full algorithmic description is given as Algorithm~\ref{alg:method1}.

\begin{algorithm}[h]
\caption{\label{alg:method1} Iris recognition, first algorithm -- case of a single query, of embedding dimension matching the CKKS ring degree}
\begin{algorithmic}
\REQUIRE{Parameters $\cN, \cP, \varepsilon_0, \Delta, \delta, \cN', \cP', \varepsilon_1$; ring degree $N$.}
\REQUIRE{$(\widehat{\MDB}, \bm^{(\db)})$ a database of $n_\db$ irises where $\widehat{\MDB}$ is a (collection of) ciphertext(s) in $\CCMM$-compatible encrypted format with ring-degree $N$ dividing~$n_\db$, along with the corresponding  masks.}\\
\REQUIRE{$(\ct^{(\qry)}_r, \bm^{(\qry)}_r)_{0\le r < \rho}$, a vector of query ciphertexts and masks.}\\
\ENSURE{Ciphertext $\ctres$ encrypting a single bit corresponding to the $\textsf{match}_{\textrm{DB}}$ function applied to the inputs.}
    \STATE $\widehat{\MQ} \leftarrow \mathsf{QueryPreprocess}((\ct^{(\qry)}_r)_{0 \le r < \rho})$
      \STATE $(\ct^{(\ell)}_{r})_{0\le r < \rho, 0 \le \ell < n_{\db}/N}\leftarrow \CCMM(\widehat{\MDB}, \widehat{\MQ})$
    \STATE $\ctres \leftarrow (0, 0)$ 

    \renewcommand{\algorithmicdo}{\textbf{do}\hfill \texttt{/* main loop */}}
\FOR{$0 \le \ell < n_{\db}/N$}
   \renewcommand{\algorithmicdo}{\textbf{do}} \FOR{$0\le r < \rho$}
   \STATE $\mu^{(\ell)}_{r} \leftarrow \Ecd\left((\|\bm_r^{(\qry)} \wedge \bm_j^{(\db)} \|_1^{-1})_{N\ell \le j < N(\ell+1)-1}\right)$ 
        \STATE $\ct^{(\ell)}_{r} \leftarrow \ct^{(\ell)}_{r} \odot \mu^{(\ell)}_{r}$
       \STATE $\ct^{(\ell)}_r \leftarrow \mathsf{Classify}_{\varepsilon_0, \cN, \cP}(\ct_r^{(l)})$
   \STATE $\ct_r^{(l)} \leftarrow \BTS(\RS_{\Delta/2^\delta}(\ct_r^{(\ell)}))$
      \STATE $\ctres \leftarrow \ctres \vee \mathsf{Classify}_{\varepsilon_1, \cN', \cP'}(\ct_r^{(\ell)})$
    \ENDFOR
    \ENDFOR\hfill \texttt{/* end of main loop */}
\renewcommand{\algorithmicdo}{\textbf{do}} \FOR{$i$ from $0$ to $\log N-1$}
    \STATE $\ctres \leftarrow \ctres \vee \Rot_{2^i}(\ctres)$
    \ENDFOR
    \RETURN $\ctres$
\end{algorithmic}
\end{algorithm}

\subsection{Query transmission and preprocessing}
A query for an eye consists of $\rho$ rotated iris codes $(\bc_r^{(\qry)})_{0 \leq r < \rho} \in \{0, 1\}^{d}$, together with the associated masks.
In the present work, we consider that only the former are encrypted, and thus restrict the discussion of packing and transmission to the iris codes.

\subsubsection{Query representation}
To reduce both communication and bootstrapping costs, we assume that the encrypted queries are sent using the smallest possible ciphertext modulus~$q_0 \approx 2^\delta$. For this purpose, we use coefficient encoding and place the plaintext in the most significant bits of ciphertexts as suggested in~\cite[Section~6]{BCKS24}.
Such a  query can be represented as $\rho$ ciphertexts $(\ct^{(\qry)}_i)_{0 \le i < \rho}$ in~$\R_{q_0,N}^2$, leading to an expansion factor of~$\approx 2 \delta$. 
This expansion factor is improved by packing bits into small integers before encrypting them. If $1 < \beta \le \rho $ is an integer packing parameter, the query can be represented as $\ct'_j = \sum_{i=0}^{\beta-1} 2^i \ct^{(\qry)}_{j\beta+i},$ for $0\le j < \lceil \rho/\beta \rceil$. This requires increasing the modulus from~$\delta$ to~$\delta + \beta$ bits, but overall reduces the expansion factor by a factor $\approx (\delta + \beta)/(\delta \beta)$.

\subsubsection{Query preprocessing}\label{sec: Query preprocessing}
The query is in a format that is unsuitable for direct homomorphic computations, its packing format being inadequate if $\beta > 1$ and, more generally, its modulus and precision being intentionally chosen very small to minimize communication costs. 
The following preprocessing steps are thus performed on the query, in order to prepare it for the linear algebra step: 
\begin{itemize}
    \item Integer bootstrapping and bit extraction~\cite{BKSS24}, in order to raise ciphertext modulus and unpack the small integers into bits; depending on the value of~$N$, this step may require applying ring-packing to the query to obtain a ring degree for which bootstrapping is possible. 
    \item Increasing the precision of the bits (see Section~\ref{sec:clean}).
    \item Conversion to the masked bit-vector representation (see Section~\ref{sec: masked bitvector}).
    Since $\bm - 2\cdot(\bc \land \bm) = \bm \odot (1-2\cdot \bc)$, this step consumes a plaintext-ciphertext multiplication.
\end{itemize}

As the cost of bit extraction quickly grows with~$\beta$, we apply the packing strategy to small values of~$\beta$. 

\subsection{Scores computation}
The core part of the iris score computation
is the evaluation of the encrypted inner products $\langle \bc_k^{(\db)}, \bc_r^{(\qry)} \rangle$ for $0 \leq k < n_{db}, 0 \leq r < \rho$.
This is viewed as the computation of a product of two encrypted matrices~$\MDB$ and~$\MQ$, for which we use the algorithm discussed in Section~\ref{sse:ccmm}.

The use of this $\CCMM$ algorithm creates a number of implementation constraints; in particular, in order to obtain the scores as ciphertexts modulo~$Q$, the database has to be represented as a ciphertext with modulus $Q^2/\Delta$. This roughly doubles the cost of the linear algebra computation, but also the storage size of the database. In order to keep both small, the natural strategy is to perform the $\CCMM$ computation at the smallest possible modulus. This produces $\rho \cdot n_\db$ scores at a low ciphertext modulus, leading to the requirement to bootstrap all these scores. For a large database, the number of ciphertexts~$\rho \cdot n_{\db}/N$ can be  high.

\subsection{Scores post-processing}
Once the scores have been computed, a post-treatment is required in order to only return relevant information, i.e., the value of $\textsf{match}_{\db}$.
The first step of this post-treatment is a classification,  implementing~\eqref{eq:condition}, and returning values in~$[0, 1]$ that are close to~$0$ (resp.~1) for scores in~$\cN$ (resp.~$\cP$). This implements the $\mathsf{match}$ function. 

Biometric inaccuracies and rotations of a match may lead to scores that are outside $\cN \cup \cP$. In this case, classification may output values strictly between~$0$ and~$1$. This is inconvenient for the OR tree. To handle this difficulty, we use a discretization step that  sends any value from~$(0,1)$ to either~$0$ or~$1$. 
Through this process, the biometrics inaccuracies may lead to false positives and false negatives, which is unavoidable given the input data. In the case of rotations of a valid query, one of the rotations is expected to output~$1$, so the others do not impact the result of the OR tree.

\subsubsection{Discretization}
 Let $(v_0, \ldots, v_{N-1}) \in [0, 1]^N$ be $N$ results of the \textsf{match} function applied to query and database irises~--~most of them are expected to be very close to~$0$ or~$1$ but some may be in-between. 
Assume a ciphertext $(a, b)$ encrypts $(v_i)_{0\le i < N}$ under~$\sk$ in coefficient encoding at modulus~$q_0$, with scaling factor~$2^\delta$, i.e.: 
\begin{align*}
    a\cdot \sk + b = \sum_{i=0}^{N-1} (2^\delta v_i + e_i) X^i \bmod q_0 \enspace,
\end{align*}
where $e_i$ accounts for the overall (computational and cryptographic) error; we assume that $|e_i| \le \ecomp$ for all~$i$.
We observe in particular that $2^\delta v_i + e_i$ is an integer in $[-q_0/2, q_0/2]$.

We half-bootstrap this ciphertext, which produces a slot-encoded ciphertext and introduces a slot-$i$ error $\varepsilon_i$ with $\max_i |\varepsilon_i| \le \ebts$, giving a value $2^\delta v_i + e_i + \varepsilon_i$. We then have the following properties: 
\begin{itemize}
    \item Values $v_i = 0$ (resp.~$v_i = 1$) are mapped to $\cN' = [-\ecomp - \ebts, \ecomp +\ebts]$ (resp. $\cP' = [2^\delta -\ecomp - \ebts, 2^\delta + \ecomp + \ebts]$); when bootstrapping precision is sufficient, $\ecomp+ \ebts$ is small compared to~$2^{\delta}$, and~$\cN'$ and~$\cP'$ are disjoint. 
    \item The slot values after half-bootstrapping lie within $\ebts$ of the integer value $2^\delta v_i + e'_i$. 
\end{itemize}
Let $n'= \lceil \max \cN'\rceil$ and $p'= \lfloor \min \cP'\rfloor$. For well-chosen parameters, we have $n' < p'$. We choose an integer $\tau \in (n',p']$ and evaluate an $([\min \cN', \tau-1+\ebts], [\tau-\ebts, \max \cP'], \varepsilon)$-classification function, obtaining a result that lies within~$\varepsilon$ of~$\{0, 1\}$, as no input to this function lies in the intermediate domain $[\tau-1+\ebts, \tau-\ebts]$. The choice of~$\tau$ is arbitrary, but can modify the probabilities of false positives and negatives. To minimize the probability of false positives, one should choose~$\tau = p'$. 

From a practical point of view, parameters should be set so as to make the classification step as efficient as possible: the relative width of the gap $(1-2\ebts)/(\max \cP' - \min \cN')$
should be as large as possible, while keeping $\ecomp + \ebts$ somewhat smaller than $2^{\delta - 1}$. This suggests that prior to this discretization, the ciphertext should be rescaled to a modulus $2^\delta$ which is as small as possible while keeping $\ecomp \ll 2^\delta$, followed by a sufficiently precise bootstrap. 

\subsubsection{Grouping everything together}
Finally, the $\textsf{match}_{\db}$ function can be obtained by evaluating an OR tree. For this purpose, we use the identity $x\vee y = x + y - x\cdot y$ in a tree-like manner with depth $\log(\rho n_{\db})$,
possibly adding cleaning steps when the precision degrades~\cite{DMPS24}. Alternatively, one may add all the $\mathsf{match}$ values and evaluate an integer indicator function to turn nonzero values into~$1$ and send~$0$ to $0$, as described in~\cite[Section~3.1]{MHPPS25}.

 \section{Reducing the number of bootstraps}
\label{se:method2}
When used for a large database, the approach of Section~\ref{sec:method1} bootstraps a very large number of scores (one per rotated iris template and database entry). Given the bootstrapping cost, this overwhelmingly dominates the overall cost (see Appendix~\ref{app:method1}). 
In particular, the bootstrapping cost exceeds by far the linear algebra cost. 
This section describes our main ingredient for rebalancing these costs, which we call \emph{folding}. It consists
of a lightweight classification (send negative scores to small values and positive scores to large ones) in such a way that we can combine several ciphertexts together before bootstrapping while retaining the relevant information. 

\subsection{Overview of folding}

\subsubsection{A high-level view}
Folding (for scalars) rests on a polynomial~$f$ such that $f(\cN) \subset [-\alpha, \alpha]$ while $f(\cP) \ge \beta$, for a small $\alpha$ compared to~$\beta$. If $(x_i)_{1\le i\le k}$ is a set of scores in $\cN \cup \cP$ such that at most one $x_i$ is not in $\cN$, we define  $S = \sum_{i=1}^k f(x_i)$ and notice that 
\begin{itemize}
    \item if all $x_i$'s are in $\cN$, then $S \le k\alpha$; 
    \item if one~$x_i$ is in $\cP$, then $S \ge \beta-(k-1)\alpha$. 
\end{itemize}

For $\alpha < \beta/(2k-1)$, we have that $k\alpha < \beta-(k-1)\alpha$, which allows us to discriminate between these two options using only the value~$S$. From the homomorphic point of view, we can thus compute~$S$, bootstrap it, and classify the resulting value to~$0$ if the input is in $[-k\alpha, k\alpha]$ and to $1$ if the input is $\ge \beta-(k-1)\alpha$. In this setting, what we compute is no longer $\textsf{match}$ nor $\textsf{match}_{\db}$ but a partial version of $\textsf{match}_{\db}$ that groups  $k$ values of $\textsf{match}$ (with an~OR) as a single one. This divides the number of bootstraps by~$k$. 

\subsubsection{Probabilistic folding}
We now make the further assumption that negative values are drawn from a probability distribution $\mathcal{D}$. We then define a folding polynomial as follows.

\begin{definition}
Given an integer $k$ and two closed intervals~$\cN_f$ and~$\cP_f$ with $\max \cN_f < \min \cP_f$, we fix two reals $p_1, p_2 \ll 1$ and define a $(k, \cN_f, \cP_f)$-folding polynomial as a polynomial~$f$ with 
\begin{align*}
\Pr_{(x_0, \dots, x_{k-1}) \leftarrow \cD^k} \left(\sum_{i=0}^{k-1} f(x_i) \not\in \cN_f\right) \le p_1\enspace, \\
\Pr_{(x_1, \dots, x_{k-1}) \leftarrow \cD^{k-1}}\left(\min_{x_0\in \cP} f(x_0) + \sum_{i=1}^{k-1} f(x_i) \not\in \cP_f\right) \le p_2 \enspace.
\end{align*}
\end{definition}

Let $\bv_0, \dots, \bv_{k-1}$ be vectors of scores of dimension $N$. We make the following \emph{folding assumption}:
for all~$0 \leq i < N$, at most one vector~$\bv_j$ has its $i$-th coordinate that is not drawn from~$\cD$.

This  simplifying assumption means that in the iris database, at most one user matches with a given query (in particular, the database contains no duplicates). We explain in Appendix~\ref{app:folding_assumption} how the following steps of our algorithmic chains can be reparameterized in order to remove it partially or totally, while almost fully preserving efficiency.

If the $\bv_j$'s satisfy the folding assumption, and~$f$ is a $(k, \cN_f, \cP_f)$-folding polynomial, we build the vector ${\bf w} = \sum_{j=0}^{k-1} f(\bv_j)$ where, for a vector~$\bu = (u_i)_i$, we let $f(\bu)$ denote the vector~$(f(u_i))_i$. Assuming that all the scores are either in $\cP$ or drawn from~$\cD$, the definition of a folding polynomial gives the following, for $0\le i < N$: 
\begin{itemize}
\item If $(\bv_j)_i$ is drawn from $\cD$ for all $j$, then $\Pr({\bf w}_i \in \cN_f) \ge 1-p_1$;
\item If there is a $j$ with $(\bv_j)_i \in \cP$, then $\Pr({\bf w}_i \in \cP_f) \ge 1-p_2$.
\end{itemize}

The result of the iris recognition process is \texttt{true} for the query index $i$ if and only if ${\bf w}_i \in \cP_f$; we deduce that, in case of a genuine match, we return \texttt{true} with probability $\ge 1-p_2$ and, in the case of a non-match, we return \texttt{false} with probability $\ge 1-p_1$. In particular, if the underlying plaintext iris recognition process has false acceptance probability $\mathsf{FA}$ (resp.\ false rejection probability~$\mathsf{FR}$), we expect to obtain a system with false acceptance probability~$\mathsf{FA'}$ (resp.\  false rejection probability~$\mathsf{FR'}$): 
\begin{align}
\mathsf{FA}' & \le \mathsf{FA} + p_1 \cdot (1 - \mathsf{FA})\enspace, \label{eq:fa}\\
\mathsf{FR}' & \le \mathsf{FR} + p_2 \cdot (1 - \mathsf{FR})\enspace. \label{eq:fr}
\end{align}

\subsection{Designing the folding polynomial}
\label{sec:folding_pol}
Given a target function $t(x)$ and a nonnegative weight function~$w(x)$, we use approximation in the $L^2$-sense,  by choosing a positive integer~$d$ and finding the degree-$d$ polynomial~$f$ minimizing
\[
\int_{x\in \mathbb{R}} w(x) \cdot (t(x) - f(x))^2 \, \textrm{d}x \enspace, 
\]
for a weight function $w(x)$ combining a probability density function~$\widetilde{\mathcal{D}}(x)$ derived from the distribution $\mathcal{D}$ that models negative scores and of the indicator function~$\mathds{1}_\cP(x)$ of~$\cP$  (see Appendix~\ref{app:folding}).

In order to design a target function, we stress that the requirements for folding are very different in nature for the negative side and for the positive side: for the negative side, the folding condition suggests that the polynomial has to approximate the zero function in the best possible way; for the positive side, however, the folding condition only requires values away from 0, but does not prescribe a precise target function. In practice, we obtained better results by allowing the function to grow slightly in the positive side. 
An example of folding polynomial is given in Figure~\ref{fig:foldP}.

Given~$w$, $t$ and~$d$, the computation of~$f$ starts by deriving an orthonormal basis of polynomials $f_0, \ldots, f_d$ of degree $\le d$ for the inner product 
\[
\langle \varphi, \psi\rangle = \int_{x\in {\mathbb R}} \varphi(x) \psi(x) w(x)\, \textrm{d}x\enspace .
\]
The folding polynomial is then obtained as $f=\sum_{i=0}^d \langle f_i, t\rangle f_i$; this reduces the problem to integration and linear algebra tasks.

\subsection{Ensuring the folding assumption} \label{sse:folding assumption}
The folding assumption asserts that in a given slot, at most one vector contains a score which is not drawn from~$\mathcal{D}$. 
Ensuring this assumption thus depends on the discrimination properties of the iris recognition system but also on the way the scores are organized among vectors.

The encrypted matrix-matrix multiplication algorithm of Section~\ref{sse:ccmm} returns a list of ciphertexts which correspond to encryptions of the columns of the underlying plaintext matrix-matrix multiplication.  As such, and in view of the specification of the iris recognition task, the output of the CCMM step is a list of ciphertexts $\ct^{(\ell)}_{r} \in \R_{Q}^2$, each containing in its $i$-th slot the score of matching the $r$-th rotation of the iris code against database entry~$i + \ell N$, where $0\le r < \rho$ and $0\le \ell < n_{\db}/N$. We let $\bv_{r}^{(\ell)}$ be the underlying vector of scores. We further assume that $\rho < N$.

\begin{figure}\includegraphics[width=1\linewidth]{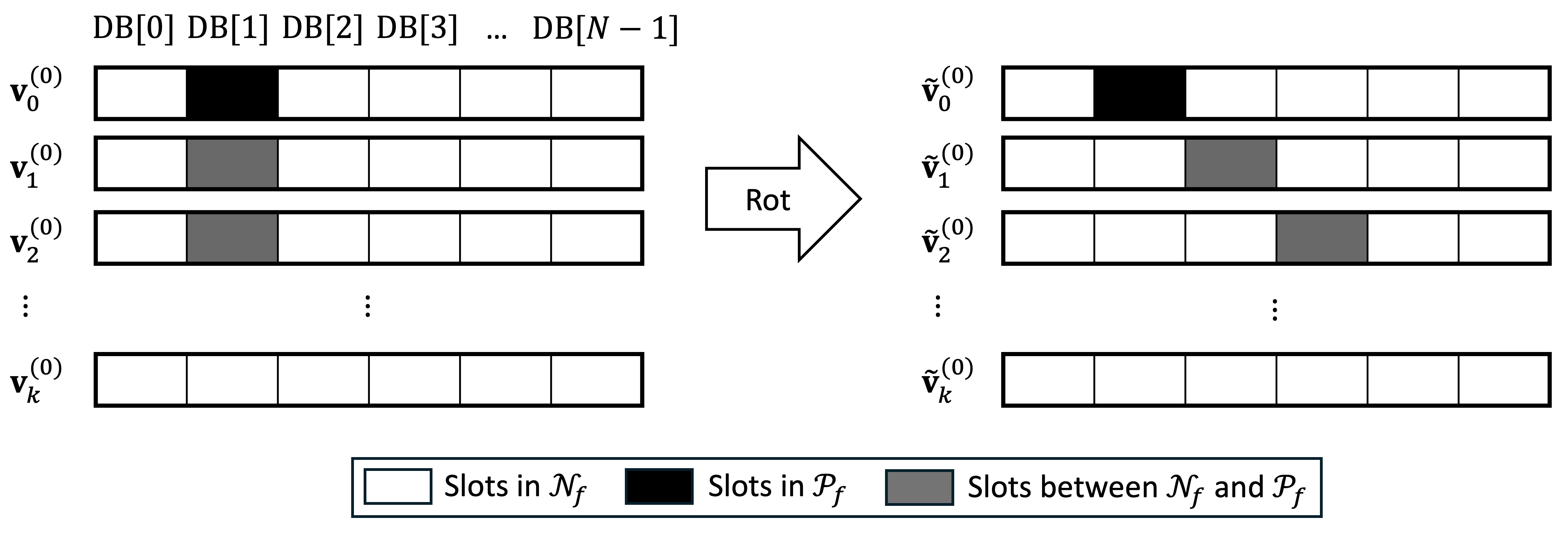}
    \caption{Among $k$ folding polynomial output ciphertexts of~$(\bv_r)_{r=1}^k$ for the same query eye, there is at most one matching DB entry. By rotating $\bv_r$ by $r$ positions, the folding assumption is fulfilled.}
    \label{fig:folding}
\end{figure}

We assume that the discrimination properties of the iris recognition system guarantee that when matching a fixed rotation of a given iris image to a set of pairwise distinct database entries (corresponding to different users), at most one of the resulting scores is not a negative one (hence not drawn from~$\cD$). 
We define $\tilde{\bv}^{(\ell)}_r = \Rot_r(\bv^{(\ell)}_{r})$;
then, the folding assumption is satisfied for the vectors $\tilde{\bv}^{(\ell)}_{r}$. Indeed, the $i$-th slot of $\tilde{\bv}^{(\ell)}_{r}$ contains a score computing against the  database entry of index $\ell N + (i - r \bmod N)$. These values are pairwise distinct for $r < N$, implying that the claim follows from our assumption on the discrimination properties of the iris recognition system. An illustration is provided in Figure~\ref{fig:folding}.

\subsection{Description of the improved algorithm}
With the above folding technique at hand, we can now describe our strategy for reducing the number of bootstraps.  We focus on the main loop of Algorithm~\ref{alg:method1}, which we improve as Algorithm~\ref{alg:method2}.

The input intervals~$\cN_f$ and~$\cP_f$ are like~$\cN'$ and~$\cP'$ in Algorithm~\ref{alg:method1}, but are set differently to take folding into account. 
The algorithm also takes as input a folding factor~$k$ and a folding polynomial~$f$.
Our folding strategy is split into two components: one just after $\CCMM$, which uses~$f$; and the second one after a first classification. At this later stage, folded values drawn from~$\cD$ have (with probability close to 1) been sent to values close to~$0$. Thanks to the folding assumption, in any given slot, at most one value is not close to~$0$; hence, we can simply add the corresponding values. Equivalently, the classification step acts as a second folding polynomial.  

\begin{algorithm}[tbp]\caption{\label{alg:method2} Improved main loop using folding}
\begin{algorithmic}
\REQUIRE{Same inputs as Algorithm~\ref{alg:method1}, and parameters~$\cN_f$ and~$\cP_f$. }
\REQUIRE{An intermediate folding amount $k \le \rho$ and a $(k, \cN_f, \cP_f)$-folding polynomial $f$.}
\STATE $(\ct^{(\ell)}_{r})_{0\le r < \rho, 0 \le \ell < n_{\db}/N}\leftarrow \CCMM(\widehat{\MDB}, \widehat{\MQ})$ \label{step:ccmm}
    \STATE $\ctres \leftarrow (0, 0)$
      \FOR{$0 \le \ell < n_{\db}/N$}
        \FOR{$0\le r < \lceil \rho/k \rceil$}
        \STATE ${\ct'_r}^{(\ell)} \leftarrow (0, 0)$
         \FOR{$0 \le r' < \min(k, \rho - rk)$}
\STATE $\mu^{(\ell)}_{rk+r'} \leftarrow \Ecd\left((\|\bm_{rk+r'}^{(\qry)} \wedge \bm_j^{(\db)} \|_1^{-1})_{N \ell \le j < N (\ell+1)-1}\right)$ 
            \STATE $\ct^{(\ell)}_{rk+r'} \leftarrow \ct^{(\ell)}_{rk+r'} \odot \mu^{(\ell)}_{rk+r'}$
            \STATE ${\ct'_r}^{(\ell)} \leftarrow {\ct'_r}^{(\ell)} + \Rot_{rk+r'}(f(\ct_{rk+r'}^{(\ell)}))$\label{step:fold}
        \ENDFOR
\STATE ${\ct'_r}^{(\ell)} \leftarrow \mathsf{Classify}_{\varepsilon_0, \cN_f, \cP_f}({\ct'_r}^{(\ell)})$
       \STATE $\ctres \leftarrow \ctres + {\ct'_r}^{(\ell)}$
       \ENDFOR
       \ENDFOR
       \STATE $\ctres \leftarrow \BTS(\RS_{\Delta/2^\delta}({\ctres}))$
       \STATE $\ctres \leftarrow \mathsf{Classify}_{\varepsilon_1, \cN', \cP'}(\ctres)$
    \end{algorithmic}
\end{algorithm}

Bootstrapping is required prior to each call to a classification function, because of the depth consumption. The total number of bootstraps in Algorithm~\ref{alg:method1} was~$\approx 2\rho n_{\db}/N$, decomposed as $\rho n_{\db}/N$ to prepare the classification of scores, $\rho n_{\db}/N$ for the discretization, and a few more for the OR tree. In Algorithm~\ref{alg:method2}, 
the number of bootstraps is~$\approx \rho n_{\db}/(kN)$ before the classification, and a few more for the discretization and the OR tree. 
Overall, the number of bootstraps is
reduced by a factor of~$\approx 2k$. 

This improvement does not come for free. The folding polynomial~$f$ needs to be evaluated on $\rho n_{\db}/N$ ciphertexts, and in order to be able to perform this evaluation before any bootstrap, the $\CCMM$ computation must take place at a higher modulus than before. Overall, bootstrap and classification  costs drop, but linear algebra becomes more expensive and a new folding cost is incurred. We stress that even though it is almost transparent from an algorithmic perspective, increasing the linear algebra ciphertext modulus is challenging in practice (see Section~\ref{sec:implem}).

 \section{System design} \label{sec:integration}

So far, we focused on the algorithmic efficiency of homomorphic matching. In this section, we explain how to deploy it in a cryptographic protocol involving encryption and threshold decryption.  

\subsection{Iris recognition using ThFHE}

We give an overview of the iris recognition system that we consider, based on threshold FHE (ThFHE). 

\subsubsection{ThFHE}
ThFHE~\cite{AJLA12,BGG+18} is an extension of FHE where the decryption key is secret-shared between~$n$ decryptors, so that at least~$t$ among them need to collaborate to decrypt and any subset of at most~$t-1$ of them cannot recover any information on plaintexts underlying ciphertexts. More formally, a ThFHE scheme consists of four protocols with the following specifications. 
\begin{itemize}
\item Key Generation is a distributed protocol between $n$ decryptors; all decryptors have scheme parameters including the security level, the number~$n$ of decryptors and a threshold~$t$ as inputs; the protocol publicly outputs an encryption key~$\pk$ and an evaluation key~$\evk$, and it outputs a decryption key share $\sk_i$ to the $i$-th party for all~$i\leq n$; 
\item Encryption is an algorithm that takes as inputs an encryption key~$\pk$ and a plaintext~$\mu$; it outputs a ciphertext~$\ct$;
\item Evaluation is an algorithm that takes as inputs an evaluation key~$\evk$, a circuit~$\mathcal{C}$ and ciphertexts~$(\ct_j)_{1 \leq j \leq k}$ where~$k$ is the number of input wires of~$\mathcal{C}$, and outputs a ciphertext~$\ct$;
\item Decryption is a distributed 1-round protocol between~$t$ decryptors; all decryptors have the same ciphertext~$\ct$ as input, the list of participating decryptors, and their own partial decryption key~$\sk_{i_j}$; the protocol publicly outputs a plaintext~$\mu$. The algorithm run by the decryptors before communication is called partial decryption: at the end of it, the~$t$ decryptors publicly broadcast decryption shares~$(\sh_{i_j})_{1 \leq j \leq t}$; these are then combined by an algorithm called final decryption which may be run by any party and produces~$\mu$.  
\end{itemize}
For efficiency purposes, we assume that all decryptors involved during decryption are aware of who the other participating decryptors are. This is typically referred to as the synchronous setting, which was studied in~\cite{mouchetthreshold,MCPT24,CPS25}. 

Correctness posits that ciphertexts produced via encryptions  and homomorphic evaluations decrypt to the plaintext evaluations of the same circuits on the plaintexts underlying the input ciphertexts. We refer to~\cite{BGG+18} for a formal definition. 

Security is defined as follows. 
The adversary can corrupt~$c<t$ decryptors.
This is assumed to occur at the outset of the game, which is referred to as selective security. Adaptive security allows the adversary to corrupt decryptors based on the ciphertexts and decryption key shares that it has viewed so far. We are not aware of any ThFHE scheme proved to be adaptively secure under standard assumptions, but: 1)~we are not aware either of any adaptive attack on a ThFHE scheme that is selectively secure and 2)~by a guessing argument, any selectively secure ThFHE scheme is adaptively secure with an~$\binom{n}{t}$ factor amplification in the adversary's security advantage. 
Beyond decryptor corruptions, the adversary can adaptively make encryption queries on valid plaintexts, evaluation queries on ciphertexts created using encryption and evaluation queries, and partial decryption queries on these ciphertexts. Such partial decryption queries are allowed for at most~$t-c-1$ decryptors for a given ciphertext. Security requires that the adversary's view can be simulated from publicly known data.   
We refer to~\cite{CPS25} for a formal definition. 

As showed in~\cite{BGG+18}, many FHE schemes (notably CKKS) can be readily converted in ThFHE schemes using the fact that decryption is essentially linear. The distributed key generation protocol from~\cite{MHPPS25} is both efficient and provides keys that are almost the same as in CKKS. In this
work, we use the extension of CKKS to a ThFHE scheme provided by~\cite{MHPPS25} for key generation and~\cite{CPS25} for decryption.

\subsubsection{Iris recognition protocol}
\begin{figure}[h]
    \centering
\includegraphics[width=\linewidth]{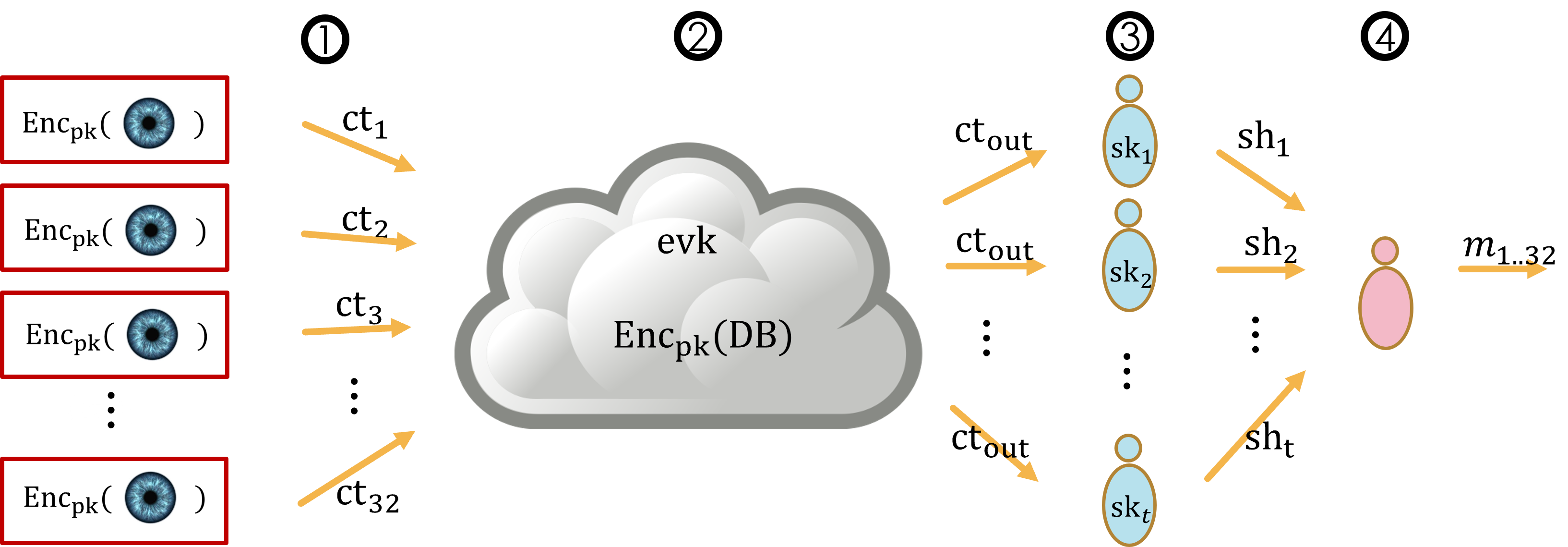}
    \caption{Batch iris recognition based on ThFHE. For the sake of simplicity, we do not depict the (plaintext) iris masks, nor potential zero-knowledge proofs required for active security.}
    \label{fig:thFHE}
\end{figure}

The computation and communication flow of the ThFHE-based iris recognition protocol is as described in Figure~\ref{fig:thFHE}. It involves four types of entities: queriers, computing servers, decryptors and a receiver.

At the outset, the decryptors run a distributed key generation: the encryption and evaluation keys~$\pk,\evk$ are made public and respectively retrieved by the queriers and computing servers; each decryptor keeps its decryption key share~$\sk_i$. Second, the database of iris codes is encrypted 
(possibly using a different encryption format than provided by the encryption algorithm) and provided to the computing servers. 

When a querier wants to match an individual against the database (Step~1 in the figure), the individual's iris codes (e.g., 31~of them per eye per user in the World ID scenario) are encrypted by the querier using~$\pk$ and the resulting ciphertext is sent to the computing servers. The computing servers wait to batch many (32 in our case) such ciphertexts to increase throughput. Using~$\evk$ and the encrypted database, they homomorphically perform the matching computation described in Section~\ref{sse:plain} (Step~2 in the figure).  The output of this computation is a ciphertext  encrypting bits (1~bit per querier) telling if there is a match or not. The corresponding ciphertext is then sent 
to the decryptors. At this point (Step~3 in the figure), each decryptor uses its decryption key share~$\sk_{i}$ and~$t$ such decryptors engage in a decryption protocol.
The decryption shares~$\sh_{i}$ are sent to the receiver. Finally (Step~4 in the figure), the receiver runs the final decryption algorithm using the decryption shares~$\sh_{i}$ to recover the plaintext output bits (1~bit per querier). 
Overall, the protocol comprises three communication rounds, and the communication is limited to the initial ciphertext, the ciphertext to be decrypted and the data sent to the receiver. More rounds and larger amounts of communication can be considered in concrete deployment scenarios to distribute computations between computing servers and assess which decryptors are available.

\subsubsection{Security model}
In the basic security model, participants are all following the protocol and try to obtain any information beyond the fact that the queries produce matches or not.
More concretely, an attacker is allowed to corrupt the computing server, the receiver and up to~$t-1$ decryptors, and to make partial decryption queries as explained in the 
ThFHE security discussion. Given this view and all ciphertexts transmitted, it should not infer any information beyond the matching result, neither on the database iris codes nor the queried iris codes. 
Also, the attacker that also corrupts the querier should not learn anything about the database iris codes beyond what it can infer from the results of the queries. Reciprocally, an attacker that knows the database iris codes (for example because it built the database) should not learn anything on the queried iris codes  beyond what it can infer from the results of the queries. Such passive security is inherited from the security of the ThFHE scheme. 

Protecting against active attackers (i.e., attackers that do not necessarily follow the protocol) can be achieved using zero-knowledge proofs. The weakest component against active attacks is decryption: indeed, the ThFHE final decryption from~\cite{BGG+18,CPS25} involves a linear combination~$b+\sum_{1\leq j \leq t} \sh_{j}$ of the decryption shares~$(\sh_{j})_{1 \leq j \leq t}$ produced by the decryptors and where~$b$ is part of the ciphertext to be decrypted; any participating decryptor can modify the output plaintext~$\mu$ to~$\mu + \mu'$ by adding (a scaled version of)~$\mu'$ to its share. 
Assuming that commitments of the decryption key shares~$\sk_i$ are made public during key generation, an approach to prevent such an attack is to have the decryptor add a zero-knowledge proof that it genuinely built~$\sh_{j}$ using its decryption key share~$\sk_{j}$. The statement to be proved is standard for lattice-based zero-knowledge proof systems (see, e.g., \cite{LNP22}).

Additionally to strengthening decryption, one could add zero-knowledge proofs inside distributed key generation to force users to follow the protocol and zero-knowledge proofs for encryption to make sure the querier is following the protocol. 
Finally, one could require efficiently verifiable zero-knowledge proofs (SNARKs) that the computing servers performed their computations correctly. Although publicly verifiable FHE schemes exist, see~\cite{CCC+25} for a very recent reference specifically for CKKS, their proving time remains prohibitive. Instead, as the computation performed by the computing servers is public, adding computation redundancy seems to be a more practical approach.

\subsection{Focus on the decryption protocol}

Assume that the computing servers have obtained a ciphertext~$\ct = (a, b)$ that decrypts to~$\mu \in \R$ that encodes bits in slots. Concretely, there exist a modulus~$q$, a scale factor~$\Delta$ and a CKKS decryption key~$\sk$ such that~$a \cdot \sk + b = \Delta \mu + e \bmod q$ where~$\|e\|_\infty$ is small compared to~$\Delta$. The decryption key is secret-shared as~$(\sk_i)_{1 \leq i \leq n}$ among the~$n$ decryptors. In partial decryption~\cite{CPS25}, the participating decryptors~$(i_1,\ldots,i_t)$ compute
\[
\sh_{i_j} = \lambda_j \cdot a \cdot \sk_{i_j} + e_{i_j} + \sum_{1\leq j' \leq t} \left(F_{k_{i_j i_{j'}}}(\ct) - F_{k_{i_{j'}i_j}}(\ct) \right)
\enspace,
\]
where~$\lambda_j$ are secret-reconstruction coefficients, $e_{i_j}$ is a fresh noise term, and~$F$ is a pseudo-random function (such as AES or using the ring version of the Learning With Rounding problem~\cite{BPR12}) and user~$j$ knows the keys~$k_{i_j i_{j'}}$ and $k_{i_{j'} i_j}$ for all~$j'$. The final decryption algorithm consists in computing
\[
b + \sum_{1 \leq j\leq t}  \sh_{i_j} = b + a\cdot \sk + \sum_{1\leq j \leq t} e_{i_j} = \Delta \mu + e + \sum_{1\leq j \leq t} e_{i_j} \enspace.
\]
One then recovers~$\mu$ from $\Delta \mu + e + \sum_j e_{i_j}$ by rounding. 

The noise term~$e_{i_j}$ hides both the computation noise~$e$ and the decryption key shares~$\sk_{i_j}$. In particular, because it is produced by homomorphic computations, the noise~$e$ may contain information on the secret key~$\sk$ and the plaintexts occurring during the computation (including the queried and database iris scores). 
If this noise~$e$ is revealed, even partially, during decryption, then attacks can be mounted~\cite{LMSS22,PS24}. To hide~$e$, the standard approach is to flood it with a much larger fresh noise~\cite{AJLA12,BGG+18}. As detailed in~\cite{LMSS22}, one can take~$e_{i_j}$ with standard deviation~$\approx 2^{\lambda/2} \cdot B_e$, where~$\lambda$ is the security parameter and~$B_e$ is a bound on~$\|e\|_\infty$. As this is to hide a single scalar, the factor~$2^{\lambda/2}$ can be multiplied by~$N$ to hide a full ring element~$e$ (applying the triangle inequality to the statistical distance). With~$N \leq 2^{16}$ and~$\lambda=128$, this amounts to~$\log N + \lambda /2 \leq 80$ extra bits of noise on top of~$e$.   
In turn, we need~$\Delta$ to be sufficiently large so that~$\mu$ can be recovered from $\Delta \mu + e + \sum_j e_{i_j}$ by rounding. Assuming that we limit the number of decryptors to~$t=2^8$, there should be a magnitude gap of~$\log N + \lambda /2 + \log t \leq 88$ bits between~$B_e$ and~$\Delta$, to 
enable secure flooding while maintaining decryption correctness. 

At the end of the homomorphic evaluation of the matching algorithm, the ciphertext may not be parameterized as above: the magnitude gap between~$B_e$ and~$\Delta$ is likely to be much smaller, e.g.,
of the order of~$4$ to~$8$ bits. To increase the gap, we homomorphically clean the message, using the technique from~\cite{KSS25}.

 \section{Implementation}\label{sec:implem}

We implemented our folding-based solution, as described in Section~\ref{se:method2},  using the HEaaN2 library~\cite{heaanlib2}.
To the best of our knowledge, HEaaN2 is the only high-performance GPU implementation of CKKS providing full flexibility of scaling factors~\cite{CCKKKMN25}. This allows to set precision according to the needs of each step, thus saving modulus and providing improved efficiency and reduced memory footprint. The solution outlined in Section~\ref{sec:method1} has not been implemented; we argue in Section~\ref{sse:results} and Appendix~\ref{app:method1} that it is bound to be at least twice slower. 

Our parameters are described in Table~\ref{tab:Param_template} (see also~\ref{tab:comp_params} in appendix)  yield 128-bit security according to the lattice estimator~\cite{APS15} and BTS failure probability below~$2^{-128}$. As indicated, we use conjugate-invariant encoding as much as possible. Some conversion steps appear in the table when they cannot be fused with other steps, and hence use one level. 

\begin{table*}[!ht]
    \centering
    \caption{CKKS Parameters. The maximum RLWE modulus is denoted~$\log PQ$; the notation $h \ (h')$ refers to the use of sparse secret encapsulation~\cite{BTH22} with general secret key Hamming weight~$h$ and temporary weight~$h'$; $\Mult$ refers to non-bootstrapping levels and the notation~``$23 \times 20$'' indicates the use $20$ levels of 23-bit scaling factor. For more details, see Table~\ref{tab:comp_params} in Appendix~\ref{app:moredetails}. \label{tab:Param_template}}
        \begin{minipage}[h]{\linewidth}
        \centering
\resizebox{1\linewidth}{!}{
        \begin{tblr}{colspec={c|c|c|c|c||c|c|c|c|c|c|c|c|c|c|c},rowsep=0pt}
            \toprule
            \SetCell[r=3]{c}{Part}
            & \SetCell[r=3]{c}{\sf Parameters}
            & \SetCell[r=3]{c}{$\log N$}
            & \SetCell[r=3]{c}{$\log PQ$}
            & \SetCell[r=3]{c}{$h~(h')$}
            & \SetCell[c=10]{c}{$\log q_i$}
            &&&&&&&&&
            & \SetCell[r=3]{c}{$\dnum$}
            \\ \cline{6-15}
            &&&&
            & \SetCell[r=2]{c}{\sf Base} & $\StC$-first
            & \SetCell[r=2]{c}{$\fromci$}
            & \SetCell[r=2]{c}{$\Mult$}
            & \SetCell[c=3]{c}{$\CtS$-first}
            &
            &
            & \SetCell[r=2]{c}{$\EvalMod$}
            & \SetCell[r=2]{c}{$\toci$}
            & \SetCell[r=2]{c}{$\CtS$}
            \\ \cline{7-7} \cline{10-12}
            &&&&&& $\StC$ &&& $\toci$ & $\StC$ & $\fromci$ &&&
            \\ \toprule
            Preprocess & Half SI-BTS
            & \SetCell[r=2]{c}{16} & \SetCell[r=2]{c}{1657} & \SetCell[r=2]{c}{1024~(32)} & \SetCell[r=2]{c}{33} & \SetCell[r=2]{c}{-} & \SetCell[r=2]{c}{20} & \SetCell[r=2]{c}{$23 \times 20$} & \SetCell[c=3]{c}{-} && & \SetCell[r=2]{c}{$52 \times 8$} & \SetCell[r=2]{c}{36} & \SetCell[r=2]{c}{$36 \times 3$} & \SetCell[r=2]{c}{3} \\ \hline
            Core & $\CtS$-first BTS
            &&&&&&&
& 20 & $20 \times 2$ & 20
            \\ \hline
            Post-process & (Half)BTS
& 16 & 1719 & 1024 (32)
& 43 & ($25 \times 2$) & (25) & $33 \times 15$
& \SetCell[c=3]{c}{-} &&
& $45 \times 8$ & 31 & $31 \times 3$
& 2
            \\
            \bottomrule
        \end{tblr}
        }
    \end{minipage}
  \end{table*}

\begin{figure}
    \centering
    \includegraphics[width=0.95\linewidth]{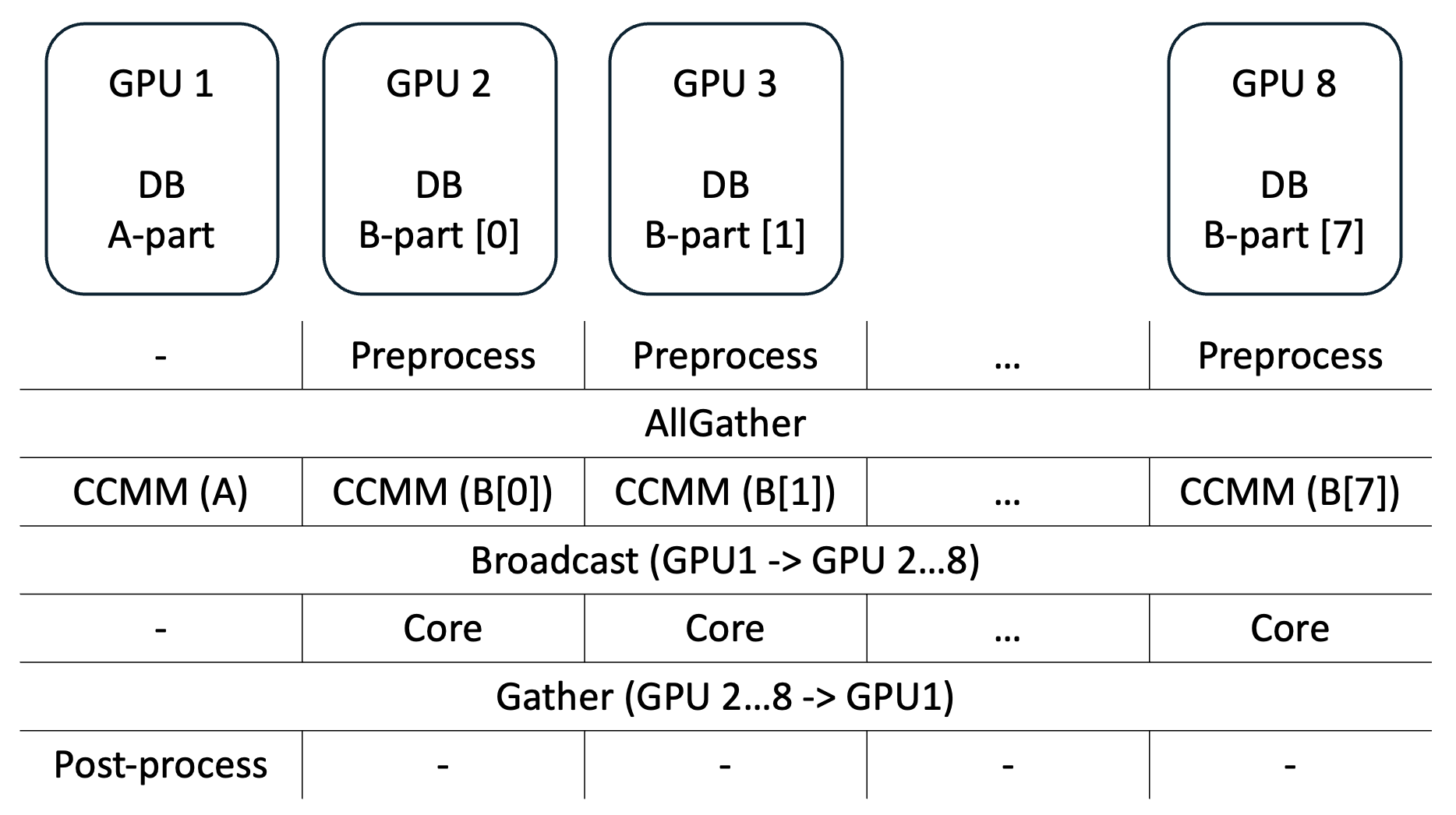}
    \caption{DB and Computation Distribution Over 8 GPUs}
    \label{fig:8GPUS}
    \vspace*{-0.5cm}
\end{figure}

Our proof-of-concept implementation considers a batch query of iris templates for 32 eyes (each one with 31 rotations), and compares them to a database of $7 \cdot 2^{14}$ iris templates. 
The database is encrypted with 8 slices. One slice corresponds to the (shared) $a$-part of 7~ciphertext slices each one of which encrypts~$2^{14}$ templates, the~7~remaining slices corresponding to the $b$-parts of the ciphertexts. In a full deployment implementation, each slice would be stored 
by 1~GPU in a cluster of 8~GPUs (see Figure~\ref{fig:8GPUS}). The run-time performance is measured on a single NVIDIA RTX-5090 GPU,  which is used to run the computation corresponding to any slice.  This gives a reliable estimate for the computation times for the full database, but not for the communication costs between the 8 GPUs. In order to estimate these, we measured the timings on a cluster of 8 RTX-4090 GPUs with PCI-e links. 

Due to space limitation, some aspects of the implementation are postponed to Appendix~\ref{app:moredetails}. Here, we focus on the linear algebra component, which is at the core of the computation and for which we considered several software and hardware accelerations.

\subsection{Dataset}
We used the ND-IRIS-0405~\cite{PSOFBSS10, NDIRIS} iris image dataset and extracted its iris templates using Worldcoin's open-source tool for plain iris recognition~\cite{openiris}.  By matching (authentic) database entries against one another, we observed that the negative scores distribution~$\cD$ can be approximated as the normal law of mean~$0.008$ and standard deviation~$0.034$.

The number of distinct eyes in the dataset is smaller than our database size, so we generated iris codes and masks  to fill the remaining database entries; the iris codes are sampled from a uniform binary distribution and masks are sampled from a Bernoulli distribution with parameter~$0.8$, assuming a moderately good image quality. 

The underlying assumption is that ND-IRIS-0405 accurately predicts concentration of rejection values around 0 and acceptance values being away from~$0$. These are the only properties that are required in the design of a folding polynomial with small~$p_1$ and~$p_2$.  
Although the actual behavior of iris templates is not simulated exactly, we are confident that this dataset extension provides a reliable estimate for the performance of our method. As shown in Equations~\eqref{eq:fa} and~\eqref{eq:fr}, the accuracy of our method is extremely close to the accuracy of the underlying biometric primitive used in~\cite{BGKSW24} as long as~$p_1$ and~$p_2$ remain  small.

\subsection{Accelerating $\CCMM$}
Encrypted matrix multiplication is used to compute the numerators of the scores.
We implemented the RGSW-based $\CCMM$ algorithm from~\cite{BCHPS25} (see Section~\ref{sse:ccmm}) using  int8 cuBLAS matrix multiplication~\cite{cublas}.

\subsubsection{RGSW-based $\CCMM$ with MSRLWE: database representation}
Following \cite{BCHPS25}, we represent the database in MSRLWE-RGSW format. We pad the iris codes to dimension $d=2^{14}$ and use ring degree $N_{\db} = 2^{14}$ for database encryption and $N'_{\qry} = 2^{13}$ for query encryption. The MSRLWE-RGSW format then consists of a pair of encryptions of $\MDB$ and $\MDB \cdot \mathbf{S}_{\qry}$, where $\mathbf{S}_{\qry}$ is a secret matrix derived from the secret keys used for the query encryption, represented as:
\begin{itemize}
    \item $2^{14}$ (resp. $2^{13}$) polynomials in $\R_{2^{14}}$ for the $a$-part of $\MDB$ (resp.\ for the $a$-part of $\MDB \cdot \mathbf{S}_{\qry}$); 
    \item $\ell \cdot 2^{14}$ (resp.\ $\ell \cdot 2^{13}$) polynomials in $\R_{2^{14}}$ for the $b$-part of~$\MDB$ (resp. for the $b$-part of $\MDB  \cdot \mathbf{S}_{\qry}$).
\end{itemize}
If~$Q$ denotes the database ciphertext modulus, the total size of the encrypted database is thus $3 (\ell + 1) 2^{27} \log Q$ bits.

\subsubsection{Fast PPMM using int8 cuBLAS} \label{sec:int8 mod square}
The RGSW-based CCMM reduces encrypted matrix multiplication to four plaintext matrix multiplications modulo $Q$. These are performed using the cuBLAS library, which offers a primitive for products of two matrices with int8 coefficients with int32 output, using the int8 tensor cores of the RTX-5090. To use this primitive, matrix multiplication modulo~$Q$ is reduced to multiplication of int8 matrices. Following~\cite{BCHPS25}, we use a combination of two techniques:
\begin{itemize}
    \item RNS representation, setting~$Q$ as a product of pairwise coprime small integers $Q=q_1 \ldots q_r$, with $q_i < 2^8$ for all~$i$; CCMM modulo~$Q$ then reduces to CCMM modulo~$q_i$, which in turn reduces to four matrix multiplications modulo~$q_i$; 
    \item Decomposition of a matrix $M$ modulo an integer $n^2$ (where $n < 2^8$) under the form $M + nM'$; then, using the identity $(M+nM')\cdot (N+nN') = MN + n(M'N+MN') \bmod n^2$, we reduce matrix multiplication modulo $n^2$ to three matrix multiplications modulo $n$.
\end{itemize}

Our target modulus is $Q \approx 2^{357}$. When using int8 base arithmetic, the RNS representation approach is limited to a total modulus of $\prod_{3\le p\le 253} p^{\lfloor 8 \log 2/\log p\rfloor} \approx 2^{354.83}$  (where the product is restricted to integers~$p$ that are prime): this falls just short. We thus use a combination of the two ideas, setting $Q = \prod_{127 \leq p \le 253} p^2$: we reduce matrix multiplication modulo~$Q$ to matrix multiplication modulo~$p^2$ for each $p$, and then  reduce each of the latter to three calls to int8 matrix multiplication. 
We note that this mixed method is more efficient than simple RNS in terms of memory representation (simple RNS requires 53 int8 matrix, compared to 48 for our solution).

\subsubsection{Database distribution and memory consumption}\label{sse:mem}
As we represent a matrix modulo $Q$ as 48 int8 matrices of the same dimension, the total size of the encrypted database is $48 \cdot 3\cdot (\ell + 1) 2^{27}$ bytes. For~$\ell = 1$, this is already $36$GB, which is larger than the main memory of an RTX-5090 GPU. 
Instead, we note that the $A$-part of the result only depends on the $A$-part of the database, while the $B$-part of the result only depends on the $B$-part of the database. We thus choose a larger value of $\ell$ and distribute the data over a computing node of 8~GPUs in the following way (see also Figure~\ref{fig:8GPUS}):
\begin{itemize}
    \item one GPU stores the whole $A$-part of the representation of the database, which accounts for $18$GB, and performs the computation of the $A$-part common to all scores; 
    \item each one of the other 7 GPUs stores the $B$-part of ciphertexts corresponding to the iris codes of a slice of the database of~$2^{14}$ users, for 18GB, and computes the corresponding $B$-part of the scores. 
\end{itemize}

This distribution allows each GPU to evaluate part of the $\CCMM$ in parallel. The $A$-part of the result, computed on the first GPU, is broadcast. The remaining GPUs can then complete the computation of the scores corresponding to their database slice.

\subsection{Experimental results}
\label{sse:results}
Table~\ref{tab:performance} shows the runtime of each step within a single GPU, assuming parallel computation over 8~GPUs. The time is measured as the average over 100 runs.
Excluding the communication between the ThFHE parties (Section~\ref{sec:integration}), it takes $\approx$1.8s to compare 32~eyes against the encrypted database of $7 \cdot 2^{14}$ iris codes. Meanwhile, the design of our folding polynomial (Section~\ref{app:folding}) leads to $p_1, p_2\approx 10^{-10}$ which, in view of $\mathsf{FA}'\le \mathsf{FA} + p_1$ and $\mathsf{FR}'\le \mathsf{FR}+p_2$ (by Equations~\eqref{eq:fa} and \eqref{eq:fr}), has negligible impact on~$\mathsf{FA}$ and~$\mathsf{FR}$.

\begin{table}[ht!]
    \centering
    \caption{Overall performance. Runtime is in ms, and the rows with a (*) symbol correspond to communication steps between the GPUs.} \label{tab:performance}
    \resizebox{\linewidth}{!}{
    \begin{tblr}{colspec={c|c|c||c|c|c},rowsep=0pt}
        \toprule
        Component & Step & Runtime & Component & Step & Runtime
        \\
        \midrule
        Core & CCMM & 259
        & Preprocess & Half SI-BTS & 136
        \\
        & Broadcast ($\ast$) & 113
        & & Extract \& Clean & 54
        \\
        & Rotate & 66
        & & FormatSwitch & 114
        \\
        & RingPack & 168
        & & AllGather ($\ast$) & 183
        \\ \cline{5-6}
        & Fold & 261
        & & Total & 487
        \\ \cline{4-6}
        & $\CtS$-first BTS & 349
        & Post-process & HalfBTS & 8
        \\
        & Classify & 89
        & & Classify & 8
        \\
        & Refold & 14
        & & Refold & 1
        \\
        & Gather ($\ast$) & 1
        & & BTS & 12
        \\
        &&
        & & Clean & 1
        \\ \cline{2-3} \cline{5-6}
        & Total & 1320
        & & Total & 30
        \\
        \bottomrule
    \end{tblr}
    }
    \vspace*{-.3cm}
\end{table}

\subsubsection*{Preprocessing}
Query preprocessing as a whole takes slightly less than~0.5s.
The ring-switching operation is relatively expensive due to the use of a very large dnum. We stress that the target ring degree is forced by the CCMM RGSW algorithm, which prevents us from increasing the ring degree to reduce dnum. 

The cost of the AllGather step could be reduced significantly by overlapping computation and communication, i.e., broadcasting parts of the queries while processing the next ones.

\subsubsection*{Core}
A notable point in this component is that the folding strategy achieves its purpose of balancing costs between linear algebra and BTS. Using the naive strategy of Section~\ref{sec:method1} has to bootstrap $16$ times more ciphertexts; multiplying the BTS cost of Table~\ref{tab:performance} already gives a computation time $> 5s$ for bootstrapping itself. We provide a more precise estimate in Appendix~\ref{app:method1}.

Most of the ring packing time comes from the encoding of the plaintext masks, which cannot be pre-computed.
The high cost of the folding step is due to its large number of inputs: 248 ciphertexts in ring degree $N = 2^{16}$, with a cost of $\approx 1.05$ ms for each.

\subsubsection*{Post-processing}
The cost for post-processing is almost negligible compared to the other components because it handles a single ciphertext.
The resulting ciphertext has 88-bit precision with a scaling factor of 105 bits over a modulus of 115 bits.

\subsubsection*{ThFHE communications}
Table~\ref{tab:performance} does not show the communication between the ThFHE parties (querier, servers, decryptors and receiver). As described in Appendix~\ref{app:more_query}, the size of a ciphertext corresponding to a user's query is 512KB (which may be lowered for a small increase of the servers' runtime). The ciphertext sent by the servers to the decryptors can be represented in ring-degree~$2^{13}$. Assuming a modulus of~128 bits (to enable flooding), this gives a ciphertext size of 256KB. In fact, only half of this (the $a$-part) is needed by the decryptors, whereas the $b$-part can be sent directly to the receiver. Each decryptor sends a decryption share of 128KB to the receiver. 
By observing that the $b$-part and decryption shares can be taken in a subring of dimension~32 (for a batch of 32 users) and truncated to a few bits, the total communication towards the receiver reduces to  less than~1KB, so that only latency matters.

 \section{Conclusion}
Our work demonstrates the relevance of ThFHE for large-scale privacy-preserving biometric recognition, offering similar efficiency as solutions based on SS-MPC while having much milder requirements in terms of communication, and allowing a greater distribution of security and computation. 

The folding technique allows one to aggressively reduce the amount of encrypted data at an early stage in large-scale FHE computations. We expect this idea to be useful in other settings. 

Finally, regarding homomorphic algorithmic design, our work illustrates major efficiency progress on homomorphic linear algebra. It becomes so efficient  
that it can be worth sacrificing some efficiency on heavy linear algebra computations (by placing them at a higher modulus) in order to reduce the cost of calls to other more expensive homomorphic primitives, such as bootstrapping.

\begin{acks}
We thank Isabelle Gu\'erin-Lassous, Jaejin Lee, Daejun Park and Minje Park for helpful discussions.
\end{acks}

\section*{AI Use}
This paper was edited for spelling and grammar using Overleaf's AI Assist. 

\bibliographystyle{alpha}
\bibliography{bibliography}

\appendix

\section{More CKKS background}
\label{app:moreCKKS}

\subsection{Bootstrapping flavors}

CKKS bootstrapping comes in a number of flavors; we use the following ones: 
\begin{itemize}
    \item the original bootstrap~\cite{CHKKS18}, which we call ``$\CtS$-first", starts with a ciphertext at the bottom modulus, and ends by the so-called "slot-to-coefficients step" ($\StC$), which homomorphically converts the encoding of the ciphertext;
    \item the ``$\StC$-first" variant moves $\StC$ from the end to the beginning of the bootstrap, and requires that the latter starts at a higher modulus;
    \item when bootstrapping a coefficient-encoded ciphertext and expecting a slot-encoded output, the StC step can simply be discarded, giving what is called ``Half bootstrapping" or HalfBTS~\cite{CHK+21}; 
    \item Dedicated solutions exist for the bootstrapping of ciphertexts encrypting vectors of small integer (SI-BTS)~\cite{BCKK25} (and Half SI-BTS for its half-bootstrapping variant).  
\end{itemize}

\subsection{Ring packing} \label{sse:ring packing}
The embedding from $\R_{Q,N} = \ZZ_Q[X]/(X^N+1)$ to $\R_{Q, 2^kN} = \ZZ_Q[Y]/(Y^{2^kN}+1)$ defined  by $X\mapsto Y^{2^k}$, can be used to map a ciphertext $(a, b)$ from~$\R_{Q, N}$ to~$\R_{Q, 2^kN}$. When using the slot encoding, in terms of the underlying message, this corresponds to sending $(x_i)_{0\le i < N/2}$ to $(x_{i\bmod N/2})_{0 \le i < 2^{k-1} N}$. From this, we notice that~$2^k$ ciphertexts $(\ct_j)_{0\le j < 2^k}$ in $R_{Q,N}$ encrypted under the same secret key can be packed to a single ciphertext in~$R_{Q', 2^kN}$ by first embedding them, and then computing
\[
\RingPack\hspace*{-.05cm}\left((\ct_j)_{j=0}^{2^k-1}\right) \hspace*{-.05cm} = \hspace*{-.1cm} \sum_{j=0}^{2^k-1} \hspace*{-.1cm}\Mult(\ct_j, \Ecd ((\underbrace{0}_{\smash{jN/2\phantom{2^k}\hspace*{-.2cm}}}, \overbrace{1}^{\smash{N/2\phantom{j2^k}\hspace*{-.3cm}}}, \underbrace{0}_{\smash{\hspace*{-.4cm}(2^k-j-1)N/2\hspace*{-.4cm}}}))) \enspace.
\]
This computation uses one multiplicative level for masking, which is realized by a plaintext-ciphertext multiplication.

\subsection{Conjugate-invariant encoding}\label{sec: Conjugate-invariant encoding}
A modified version of CKKS uses the \emph{conjugate-invariant} subring $\R_{\ci}=\ZZ[2\cos(2\pi/N)]$ as plaintext space and~$\RR^N$ as message space~\cite{KS18}: this provides~$N$ real-valued slots in contexts where one has no use for complex numbers. As our computations handle real numbers, we only use conjugate-invariant encoding. Some homomorphic primitives (bootstrapping, ring-switching), however, are more efficient using complex encoding. The  conversions can be performed without decrypting~\cite{BCKK25} and will be denoted by $\toci$ and~$\fromci$. They are cheap but consume one multiplicative level.

\section{Classification functions}\label{app:classification}
Several techniques have been developed for building classification functions, mostly in the context of evaluating comparisons. Two families of approaches can be considered.
\begin{itemize}
    \item Use a composition of small degree polynomials; this consumes fewer homomorphic operations, but suffers from higher multiplicative depth and hence often more bootstraps~\cite{CKK20};
    \item Use a large degree polynomial constructed using the multi-interval Remez algorithm, and evaluated using the Paterson-Stockmayer algorithm~\cite{PaSt73}. This decreases multiplicative depth, but may lead to very large degrees and hence high costs. 
\end{itemize}

These two methods are the two ends of a spectrum corresponding to a tradeoff between depth (or modulus consumption) and computational cost; as such, intermediate solutions have been considered, by composing medium-degree polynomials built using the multi-interval Remez algorithm~\cite{LLNK22, LLKN22b}; we shall follow this approach. In order to build an $(I_0,I_1, \varepsilon)$-classification function given a bound~$d$, we use the multi-interval Remez algorithm to find a minimal~$\varepsilon_0$ and a degree-$d$ polynomial~$P_0$ such that~$P_0$ is a $(I_0, I_1, \varepsilon_0)$-classification function. Then, for $Q \circ P_0$ to be an $(I_0, I_1, \varepsilon)$-classification function, it suffices that~$Q$ is an $([-\varepsilon_0, \varepsilon_0], [1-\varepsilon_0, 1+\varepsilon_0], \varepsilon)$ classification function; we then use the same idea recursively, obtaining a sequence of~$\varepsilon_j$'s, until $\varepsilon_j \le \varepsilon$.

The cost of classification in CKKS is related both to the relative width of the gap between $I_0$ and $I_1$, namely 
\[
\frac{\min(I_1)-\max(I_0)}{\max(I_1) - \min(I_0)}
\]
and, in a more subtle manner, to the position of the gap (classification is easier if one of the intervals $I_0$ or $I_1$ is much smaller than the other one).

\section{More details on the implementation}
\label{app:moredetails}

\begin{figure*}
    \centering
    \includegraphics[width=0.9\linewidth]{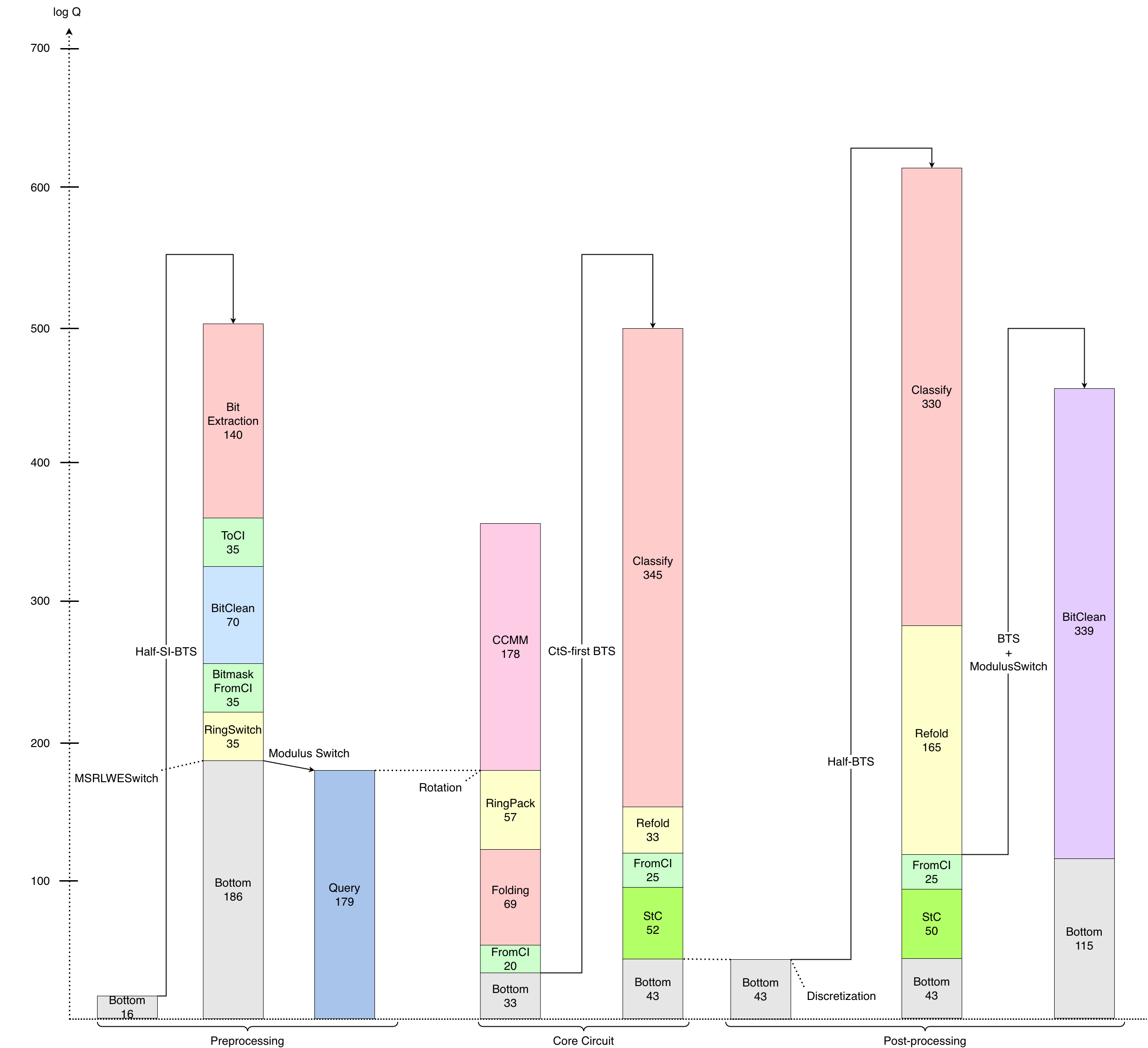}
    \caption{Modulus allocation} \label{fig:modulus}
\end{figure*}

The detailed modulus consumption of the various steps is provided in Figure~\ref{fig:modulus}, while complete FHE parameters are given in Table~\ref{tab:comp_params}.
    \begin{table*}[h]
    \centering
        \caption{Computing Parameters.\label{tab:comp_params} }
        \resizebox{1\linewidth}{!}{
        \begin{tblr}{colspec={c|c|c|c|c||c|c|c|c|c}}
            \toprule
            \SetCell[r=2]{c}{Part}
            & \SetCell[r=2]{c}{Parameters}
            & \SetCell[r=2]{c}{$\log N$}
            & \SetCell[r=2]{c}{$\log PQ$}
            & \SetCell[r=2]{c}{$h$}
            & \SetCell[c=4]{c}{$\log q_i$} &&&
            & \SetCell[r=2]{c}{$\dnum$}
            \\ \cline{6-9}
            &&&&& {\sf Base} & $\StC$ & $\fromci$ & $\Mult$
            \\ \hline
Preprocess
            & Bit Extract \& Clean
& 16 & 1050 & 1024
& \SetCell[c=4]{c} $186 + 35 \times 9\text{ [masking + fromCI + clean + toCI + BitExtract]}$ &&&
& 1
            \\ \cline{2-11}
            & Format Switch
& 13 & 212 & 512
& \SetCell[c=4]{c} $23\times 6 + 24\times 2 $ &&&
& 8
            \\ \hline
Core
            & Query Encrypt
& 13 & 179 & 512
& 33 & - & 20 & $23 \times 3\text{ [fold]} + 57\text{ [ring pack]}$
& -
            \\ \cline{2-11}
            & DB Encrypt
& 14 & 357 & 512
& 33 & - & 20 & $23 \times 3\text{ [fold]} + 57\text{ [ring pack]} + 178\text{ [CCMM]}$
& -
            \\ \cline{2-11}
            & RingPack \& Fold
& 16 & 251 & 32
& 33 & - & 20 & $23 \times 3\text{ [fold]}$
& 1
            \\ \cline{2-11}
            & Classify \& Refold
& 16 & 1058 & 1024
& 43 & $26 \times 2$ & 25 & $33\text{ [refold]} + (35 \times 3 + 30 \times 4 + 30 \times 4)\text{ [classify]}$
& 1
            \\ \hline
Post-process
            & Classify \& Refold
& 16 & 1284 & 1024
& 43 & $25 \times 2$ & 25 & $(33 \times 5)\text{ [refold]} + (33 \times 5 + 33 \times 5)\text{ [classify]}$
& 1
            \\ \cline{2-11}
            & Clean
& 16 & 1058 & 1024
& \SetCell[c=4]{c}{$115\text{ [output]} + (123 + 92 + 52 + 72)\text{ [clean]}$} &&&
& 1
            \\
            \bottomrule
        \end{tblr}
        }
    \end{table*}

\subsection{Query preprocessing}
\label{app:more_query}
Each query is initially packed in vectors with coordinates that are integers of $\beta = 4$ bits. They are encrypted using a low ciphertext modulus~$q_\pre = 2^{16}$ to optimize communication and BTS. We use the RLWE format with ring degree $N = 2^{16}$ (the MSRLWE format with a smaller ring degree could be used to further reduce communication, at the expense of a ring-packing cost on the server side). For a query dimension~$d \le 2^{14}$ and $\rho = 31$ rotations, each rotation is encrypted as a pair of ciphertexts, for a total size of $512$KB. The whole input thus consists of 62 ciphertexts, distributed over the 7 GPUs of our cluster. Each GPU processes its part of the query, all parts being broadcast after preprocessing in the AllGather step of Figure~\ref{fig:8GPUS}. 

We use Half-SI-BTS followed by bit extraction as described in~\cite{BKSS24}, using homomorphic parameters described in~Table~\ref{tab:Param_template} part (a) for bootstrapping and~part (b) for bit extraction.
We then clean the ciphertexts to recover enough precision, and turn them into the bitmask format~(see Section~\ref{sec: masked bitvector}) via
\[
    \ct_\mathsf{mask} = \Mult(1 - 2\cdot \ct, m) \enspace. 
\]

To prepare for CCMM, these ciphertexts are ring-switched to ring-degree $2^{13}$, each encrypting half a masked iris code. Each pair of half iris codes is converted to MSRLWE using format-conversion as in~\cite{BCHPS24}, and the scaling factor is reduced to the correct one for CCMM by modulus switching. In Table~\ref{tab:performance}, we group these steps under the label ``FormatSwitch''. 

Ring-switching is performed in slots encoding. For this, we use masking to extract partial ciphertexts encrypting $2^{13}$ consecutive slots, then compute the trace (or, equivalently, partial rotate-and-sum) of each of these partial ciphertexts to bring the ciphertext down to the ring $\R_{2^{13}}$. 
The traces can be evaluated efficiently by first key-switching the partial ciphertext to a secret key in $\R_{2^{13}}$, and noticing that taking the trace of the ciphertext is then equivalent to taking the traces of the~$a$-part and~$b$-part, which corresponds to extraction of coefficients of degrees divisible by~$8$.

\subsection{Folding and bootstrapping}\label{app:folding}

This part of the computation chain needs to handle a strong modulus constraint: the modulus~$Q$ used between CCMM and the modulus 
raising step of bootstrapping impact the total modulus~$Q^2/\Delta$
used for the database representation and, hence, the total database 
storage, which is a critical resource in our implementation. 

Using the strategy of Section~\ref{sec:folding_pol}, we compute a folding polynomial $P$ of degree 7 with folding capability $k = 16$. The evaluation of this polynomial requires 3 levels, for which we use a scaling factor of $\Delta = 2^{23}$. In order to retain as much precision as possible under this small scaling factor, we use secret key Hamming weight $h = 32$ for this computation.

In order to construct the folding polynomial, we consider $w_{\alpha}(x) = \alpha \mathcal{D}(x) + \mathds{1}_\cP(x)$, as a weight function, with $\alpha = 10^3\gg 1$ to put more emphasis on constraining the behavior on the negative scores than on the positive scores. This unbalancedness is motivated by the addition of many negative scores.
For the distribution~$\widetilde{\mathcal{D}}$, we considered the normal law with mean~$0.008$ and standard deviation~$0.06$. 
As a target function, we used $x \mapsto (2+x) \cdot \mathds{1}_{\cP}(x)$ (where~$\mathds{1}_{\cP}$ denotes the indicator function of~$\cP$). 
All these parameters have been experimentally optimized. 

\begin{center}
\begin{figure}
\caption{\label{fig:foldP}Graph of the 16$\times$ folding polynomial}
\includegraphics[width=0.9\linewidth]{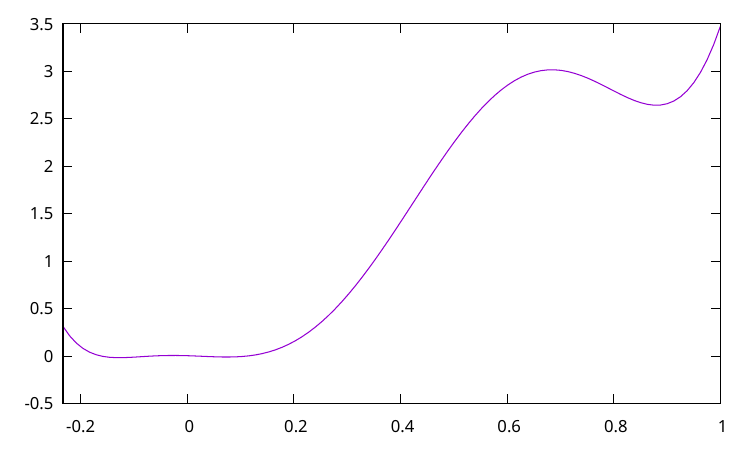}
\end{figure}
\end{center}

Folding operates on $16$ values drawn from~${\mathcal D}$ at once, leading 
to a larger occurrence of outlier values; for the design of the polynomial, using a distribution~$\widetilde{\mathcal{D}}$ with larger standard deviation simulates this larger occurrence of outliers. We obtain the polynomial
\begin{align*}
f(x) & = 106.553952\cdot x^7 + 376.961251\cdot x^6 - 412.746212\cdot x^5\\
& + 124.161550 \cdot x^4 + 24.347349 \cdot x^3 \\ & - 
2.528271\cdot x^2 - 0.173510\cdot x + 0.004105 \enspace.
\end{align*}
Figure~\ref{fig:foldP} displays the graph of~$f$ over~$[-0.233, 1]$. We see that it takes values that are consistently very close to~$0$ around~$0$, and that large inputs have large but non-constant values. 

In order to validate our folding design, we ran $10^{12}$ folding experiments computing $\sum_{0 \leq i < 16} P(X_i)$, $\min_{x\in \cP}P(x) + \sum_{0 \leq i \leq  14} P(Y_i)$ and~$\max_{x\in \cP}P(x) + \sum_{0 \leq i \leq 14} P(Y_i)$ with $X_i, Y_i$ independently drawn from~$\cD$. We found:
\begin{itemize}
    \item 53 values coming from the sum of 16 values drawn from~$\cD$ outside of $[-0.13, 0.33]$; 
    \item 542 values coming from the sum of 15 values drawn from~$\cD$ and one value in $\cP$ outside of $[0.4, 3.8]$. 
\end{itemize}

We take $\cN_f = [-0.13, 0.33]$ and $\cP_f = [0.4, 3.8]$ in Algorithm~\ref{alg:method2}, and heuristically estimate $p_2 \approx 5.3\cdot 10^{-11}$ and  $p_1 \approx 5.4\cdot 10^{-10}$. It implies, in view of Equations~\eqref{eq:fa} and~\eqref{eq:fr}, that folding should have no observable impact on $\mathsf{FA}$ and~$\mathsf{FR}$. In order to account for the homomorphic evaluation error, we replace~$\cN_f$ by~$\cN_f' = [-0.15, 0.35]$. 

After evaluating the folding polynomial, we add the resulting ciphertexts, ring-pack to $\log N = 16$ and bootstrap. This bootstrap is performed using the CtS-first variant, which starts by raising the modulus. The StC-first variant would lead to an increased database memory footprint.

\subsection{Classification and refolding}
The classification steps are implemented using the general strategy outlined in~Section~\ref{sec:clean}; the input and output intervals for the classification step, and the degrees of the intermediate polynomials can be found in Table~\ref{tab:classify}.
The various scaling factors can be found in Table~\ref{tab:Param_template}.

\subsubsection{Refolding}
The first folding folds only~$16$ values, even though the folding assumption holds for~$2^{14}$ values. To decrease the number of ciphertexts to be manipulated and communicated during the rest of the computation, we refold whenever possible to decrease the number of ciphertexts to a single one. 

\subsubsection{Gathering}
In our experiment, 7 GPUs perform the folding computation and classification part of the algorithm; on a $7\cdot 2^{14}$ database for a batched query of 31 rotations of a 32 different iris images, each GPU handles 248 ciphertexts prior to folding, a number reduced to 16 after folding. 

In order to complete the folding after the first classification, these ciphertexts are sent to a single GPU, the latter handling the rest of the computation. In order to minimize the communication cost, we perform the StC part of the following bootstrap before gathering the ciphertexts, so that only HalfBTS has to be performed afterwards. The resulting communication costs appear as ``Gather" in Table~\ref{tab:performance}.

\begin{table}
    \centering
    \caption{Classification functions}
    \label{tab:classify}
    \resizebox{1\linewidth}{!}{
    \begin{tblr}{c|c|c|c|c}
        \toprule
        \SetCell[r=2]{c}{Part}
        & \SetCell[c=2]{c}{Input Intervals} &
        & \SetCell[c=2]{c}{Polynomials} &
        \\ \cline{2-5}
        & $\cN$ & $\cP$ & Degree & Scale Factors
        \\
        \midrule
        \SetCell[r=3]{c}{Core}
        & \SetCell[r=3]{c}{$[-0.15, 0.35]$} & \SetCell[r=3]{c}{$[0.4, 3.8]$}
        & 15 & 30 bits
        \\
        &&& 15 & 30 bits
        \\
        &&& 7 & 35 bits
        \\
        \midrule
        \SetCell[r=2]{c}{Post-processing}
        & \SetCell[r=2]{c}{$[-0.414, 0.571]$} & \SetCell[r=2]{c}{$[0.585, 1.414]$}
        & 31 & 36 bits
        \\
        &&& 31 & 36 bits
        \\
        \bottomrule
    \end{tblr}
    }
\end{table}

\subsubsection{Post-processing}
Post-processing outputs binary results that are packed into a single bit per query eye, with high enough precision to enable noise flooding (which is itself necessary for ThFHE security).
The first step of the post-processing is a discretization to handle non-binary results.
Given a coefficient-encoded input ciphertext, we rescale it to reduce its scaling factor to $2^5$ (i.e., we set $\delta = 5$), and bootstrap.
This creates a gap between~$0.571$ and~$0.585$, which we use to classify as mentioned in Table~\ref{tab:classify}. We conclude with high-precision cleaning as in~\cite{KSS25}.

\subsection{On the folding assumption}\label{app:folding_assumption}
The folding assumption (see Section~\ref{sse:folding assumption}) states that 
\begin{quote}
for all~$0 \leq i < N$, at most one vector~$\bv_j$ has its $i$-th coordinate that is not drawn from~$\cD$.
\end{quote}

Its main purpose is to simplify the discussion of the output of folding, and the subsequent refolding steps: after folding, slot number $i$ contains either $\sum_{j=0}^{k-1} f(v_{j,i})$, where all $v_{j,i}$ are drawn from $\cD$, or \[
f(v_{j,i_0}) + \sum_{\substack{j=0 \\ j\neq i_0}}^{k-1} f(v_{j,i}),
\]
where $v_{j,i_0} \in \cP$, the other $v_{j,i}$ being in $\cD$; in the first case, the values end up in $\cN_f$ with high probability, while they end up in $\cP_f$ in the second case. 

We outline how one can remove the folding assumption. Assuming that for a fixed $i$ we may have up to $p$ elements of $\cP$ among $\{ v_{j,i}, 0\le j < k-1\}$, and given that $0\in \cN_f$, we deduce that the sum $\sum_{j=0}^{k-1} f(v_{j,i})$ is, with large probability, 
\begin{itemize}
    \item either in $\cN_f$ (all negatives)
    \item or in $\cP_{f,p} + [(p-1)\min f(\cP), (p-1)\max(\cP)]$ (at least one positive), 
\end{itemize}
where, for two sets $A, B$ we write $A+B = \{a+b, a\in A, b\in B\}$.

Assuming that $\min f(\cP)>0$, we set $\cN'_f = \cN_f$ and 
\[\cP'_f = [\min(\cP_f), \max(\cP_f)+(p-1)\max f(\cP)],\]
and claim that $\cP'_f$ contains all folded values coming from sets with at most $p$ positive values. Modifying the folding assumption to "at most $p$ positive values" amounts thus to turn a $(\cN_f, \cP_f)$ classification problem to a $(\cN'_f, \cP'_f, \varepsilon)$ classification problem. 

This classification problem is harder, as the gap between $\cN'_f$ and $\cP'_f$ divided by the size of the convex hull of $\cN'_f \bigcup \cP'_f$ decreases; however, the gap moves closer to the border of the latter interval, which has the effect to make the problem easier at the same time. 

We tested complete removal of the folding assumption in the setting of our experiments of Section~\ref{sec:implem}. We obtain $\cN'_f = [-0.15, 0.35]$ while $\cP'_f = [0.4, 3.8+15 \cdot 3.5]$. Increasing the degree of the last polynomial from 7 to 15 then leads to a classification of sufficient accuracy for the following refolding, at a slight increase of cost. 

We note that in the absence of the folding assumption, the following refolding steps may also produce values $> 1$; similar modifications apply in that setting. 

\subsection{Scaling estimates}
We discuss the scaling properties of our approach for a larger database, or for a larger or smaller query batch size. 

\subsubsection{Handling a larger database}
In our experiments, database storage makes almost full use of the available GPU memory. Scaling up thus either requires increasing the amount of available hardware or move to hardware with other characteristics.
For instance, running the experiment of~\cite{BGKSW24}, which handles a database of size  $2^{22}$, would require 37 GPU clusters of the same type as the one that we consider (8~\mbox{RTX-5090} GPUs). The cost of this configuration is comparable to that of the 24 H100 GPUs used in~\cite{BGKSW24}.

Scaling while keeping the same hardware would mean loading data from CPU memory after completing the CCMM step. This does not seem realistic: the next slice of $7 \times 2^{14}$ entries of the database has size $\approx 144$GB and would require 5~to 7~seconds to transfer from main memory using the PCI bus. This is too large to significantly overlap with computation. In order to explore this direction, other hardware configurations have to be considered. 

\subsubsection{Handling a larger batch size}
Handling larger query batch sizes is a different matter. For our batch of 32 query eyes, query size remains moderate and it is possible to increase it slightly within the available GPU RAM. For larger batch sizes, there is thus a tradeoff between scaling sequentially (by processing a large batch as several sequential smaller batches) or in parallel (by increasing the hardware configuration). For instance, the authors of~\cite{BGKSW24} consider a batch of 32~users, but with 2~eyes per user. Compared to our experiments, this requires doubling the query batch size, which we can handle in twice more time or in the same time by doubling the number of GPUs. 

\subsubsection{Handling a smaller batch size}
Scaling down the query batch size does not allow a reduction in hardware configuration (because of the database storage), but yields faster processing time, at least down to 2 query eyes (which corresponds to handling a single ciphertext per GPU after folding). For hardware reasons (smaller queries change the balance between compute-bound steps and memory-bound steps), we do not expect the scaling to be fully linear. For instance, we ran the core circuit on an $8\times$ smaller batch of 4 query eyes, and obtained a total time of $200$ms for computational steps, so only a reduction of $6\times$; for such a batch, the end-to-end timing is expected to be of the order of $0.3$s. 

\subsection{Estimated cost of the first approach}\label{app:method1}
We did not implement the first algorithm because its bootstrap cost alone already gives a conservative lower bound that exceeds the running time of our folded implementation. To make this comparison concrete, consider a database of size $8 \times 2^{16}$, distributed over $8$ GPUs with each GPUs storing $2^{16}$ entries. In the first approach, the scores are bootstrapped before classification, thus the number of bootstraps required is at least $32 \times 31$, where $32$ is the query batch size and $31$ is the number of rotations.
Using an $10$ms latency per bootstrap on an ~\mbox{RTX-5090}, this already gives the following per-GPU running-time lower bound 
\[
    32 \cdot 31 \cdot 10 \mathrm{ms} \approx 9.9 \mathrm{s} \enspace.
\] 

For comparison, the core part of our folded approach takes $1.3$s for a database of size $7 \cdot 2^{14}$ entries. Linearly scaling this to $ 8 \cdot 2^{16}$ database entries gives:
\[
    \frac{8}{7} \cdot \frac{2^{16}}{2^{14}} \cdot 1.3 \mathrm{s} \approx 5.9 \mathrm{s} \enspace.
\]
Therefore, the folded approach is estimated to be at least $9.9/5.9 \approx 1.7$ times faster. 

Nevertheless, the memory requirements of our first method are lower. As such it can provide a solution which, though much less efficient, is easier to deploy, requiring fewer GPUs overall, at the cost of performance loss.

\end{document}